\documentclass[%
reprint,
 amsmath,amssymb,
 aps,
 prb,
]{revtex4-1}

\usepackage{graphicx}
\usepackage{caption} 
\usepackage{subfig}
\usepackage{dcolumn}% Align table columns on decimal point
\usepackage{bm}% bold math

\graphicspath{{image2/}}

\begin{document}

%\preprint{}

\title{Spin Transport Properties of Fractal and Non-Fractal Thinfilm}


\author{Cheng-Yen Ho}
\email[Electronic mail: ]{r97222052@ntu.edu.tw}
%\thanks{}
%\affiliation{Department of Physics, National Taiwan University, Taipei 10617, Taiwan}
\author{Ching-Ray Chang}
\email[Electronic mail: ]{crchang@phys.ntu.edu.tw}
%\thanks{}
\affiliation{Department of Physics, National Taiwan University, Taipei 10617, Taiwan}

\date{\today}

\begin{abstract}
Spatial behavior of spin transport in a Sierpinski gasket fractal is studied from two dimensions to quasi-one dimension subject to the Rashba spin-orbital coupling. With two normal metal leads represented by self-energy matrix, discretizing the derived continuous  Hamiltonian to a tight-binding version, Landauer-Keldysh formalism for nonequilibrium transport can be applied. It was observed that the spin Hall effect presents in the distribution of spin density critically depends on the fractal structure and the shape of the thinfilm. The local spin density and transmission are numerically tested by the present quantum transport calculation for the fractal and non-fractal thinfilm varying from two-dimensional square-lattice into quasi-one dimensional fractal shape (Sierpinski triangles).

\begin{description}
\item[PACS numbers]
{72.25.–b,73.63.Nm,71.70.Ej}
\end{description}

\end{abstract}

%PACS numbers: 72.25.–b,73.63.Nm,71.70.Ej
%\keywords{}

\maketitle

%\ section{}
%%\ section{\ label{}}
%\ subsection{}
%\ subsubsection{}

\section{Introduction}

Intensive efforts on spin Hall effect (SHE) both experimentally and theoretically have successfully built another milestone in condensed matter physics\cite{PhysRevB.71.195314}. Spin separation in semiconductors is not only possible but natural, so that manipulating spin properties of charge carriers in electronics is promising\cite{Lodder1999119,PhysRevLett.92.086601,PhysRevLett.93.226602}. Recently different types of SHE were reported on different geometrical setup of both magnetic\cite{Sun:2010zr,Sun:2010ly} and non magnetic materials\cite{PhysRevLett.98.156601,PhysRevLett.106.157208}. Local spin Hall effect often attributed as extrinsic SHE\cite{Kato10122004,Sih:2005fk,Seki:2008uq}, and intrinsic SHE\cite{PhysRevLett.94.047204}. The intrinsic SHE, spin separation is solely due to the underlying spin-orbit coupling in the band structure, so that SHE can exist even in systems free of scattering within a finite size system\cite{PhysRevB.70.041303,PhysRevB.82.155327}. However, experimentally, most observations so far have been attributed to the extrinsic SHE. More recently nonlocal spin Hall effect was also found in various systems\cite{Sun:2010zr,Seki:2008uq,Roth17072009}, including graphene\cite{Sun:2010ly,Abanin15042011,PhysRevB.78.165316}. Considerable effort in this direction has already revealed the unique features of spin-polarized electron transport in the so-called two-terminal nano- or mesoscopic devices. Indeed, this is a fast developing field, and has also stimulated lot of theoretical work based on the non-equilibrium Green's function approach within the density functional theory \cite{PhysRevB.78.165316,PhysRevLett.95.046601,Datta:1998fk}. 

A two-terminal device is essentially a single path device, SHE studies on a two-terminal device with defects show a lot interesting results and the local spin densities can be affected seriously even by a small point defects\cite{chen:07E908,Chen:qy} for the breaking of translational symmetry. In the present studies we undertake an in-depth study of the spin transport in a fractal network based on the non-equilibrium Green's function approach, the Landauer-Keldysh formalism (LKF)\cite{PhysRevB.78.165316,PhysRevLett.95.046601,Datta:1998fk}. We choose a fractal geometry following a Sierpinski gasket (SPG) \cite{0295-5075-10-1-013,PhysRevB.28.3110} and examples of planar gasket is shown in Fig. \ref{fig:fc}. With the self-similarity of the system and the enlargement of the fractal structure, our studies provide an understanding of the spin transport behavior transition from two dimensional system to quasi-one dimensional system. Our motivation for the investigation of spin transport within fractal system is mainly resulted from our previous studies on impurity system. The defect and impurities serious affect the local charge density and local spin density for the influences on spin transport, moreover, the relations between spin current and charge current within a film with defect behave very different from that within an ideal film\cite{Chen:qy,chen:07E908}. Even though the point-defect induced translational symmetry breaking and interference among different conducting channels provide subtle reasons for the variation of the spin density and current density, the detailed analysis of defect and impurities of the multi-terminal systems is very rare and the multiply connected fractal geometry is not even been probed. The state-of-art lithographic technologies made all artificial structures become possible. Therefore, the studies of spin transport of self-similarity of systems such as the SPG provide the possibility of investigating the local density of charge and spin on different multiply connected fractal geometry. 

Secondly, a very unique phenonemon in spin transport remains to be understood is the spin processional phase through different travelling channel, therefore, the quantum interference among the confined paths of the fractal structures will be an interesting and provide a good way of studying the self-similarity with reducing scale. Even though the charge transport on percolating clusters had been studied but  the local charge density and local spin density, and the Aharonov-Bohm effect in a fractal system without translational invariance remains unclear till now\cite{PhysRevLett.52.1033,PhysRevLett.57.649}. Moreover, the density of transistors on an integrated circuit doubles every couple of years. The scale of devices is one of important factors. Fractal structures also can study the limits of the self-similarity with reducing scale for a profound 2D-system.

The paper is organized as follows, we present Landauer-Keldysh formalism with a tight-binding framework to model the fractal system in Section II. Numerical results of local charge density and local spin density of different size of fractal system are studied and comparison among continuous film in low dimensional system also been discussed in Section III. We draw our conclusions in Section V.
%{As Moor's low, the density of transistors on an integrated circuit doubles every couple of years. At first, electronic industries focus on the frequency of devices. And now, they reduce the number of devices in the same area. The scale of devices is one of important factors. Fractal structures provide a good way of considering reducing scale.\\}

\section{Non-equilibrium Green function in Sierpinski gasket}

\subsection{The Sierpinski gasket}
\begin{figure}[ht]

\centering
\subfloat[First generation of SPG.]{
\label{First generation of SPG.}
\begin{minipage}[h]{0.22\textwidth}
	%\centering
	\includegraphics[width=1.7in]{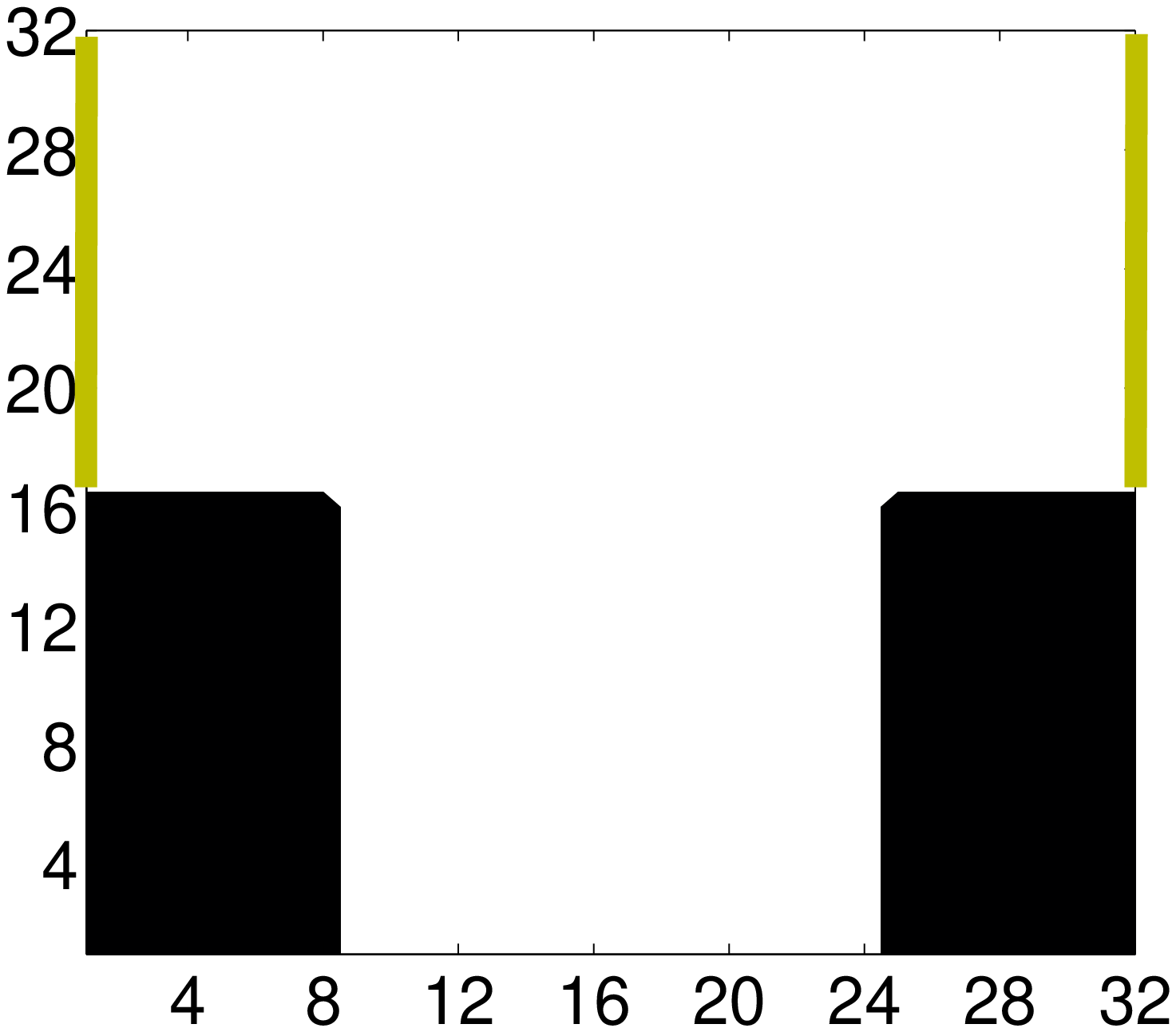}
	%\caption{Case 1}	
\end{minipage}
}
\subfloat[Second generation of SPG.]{
\label{Second generation of SPG.}
\begin{minipage}[h]{0.22\textwidth}
	%\centering
	\includegraphics[width=1.7in]{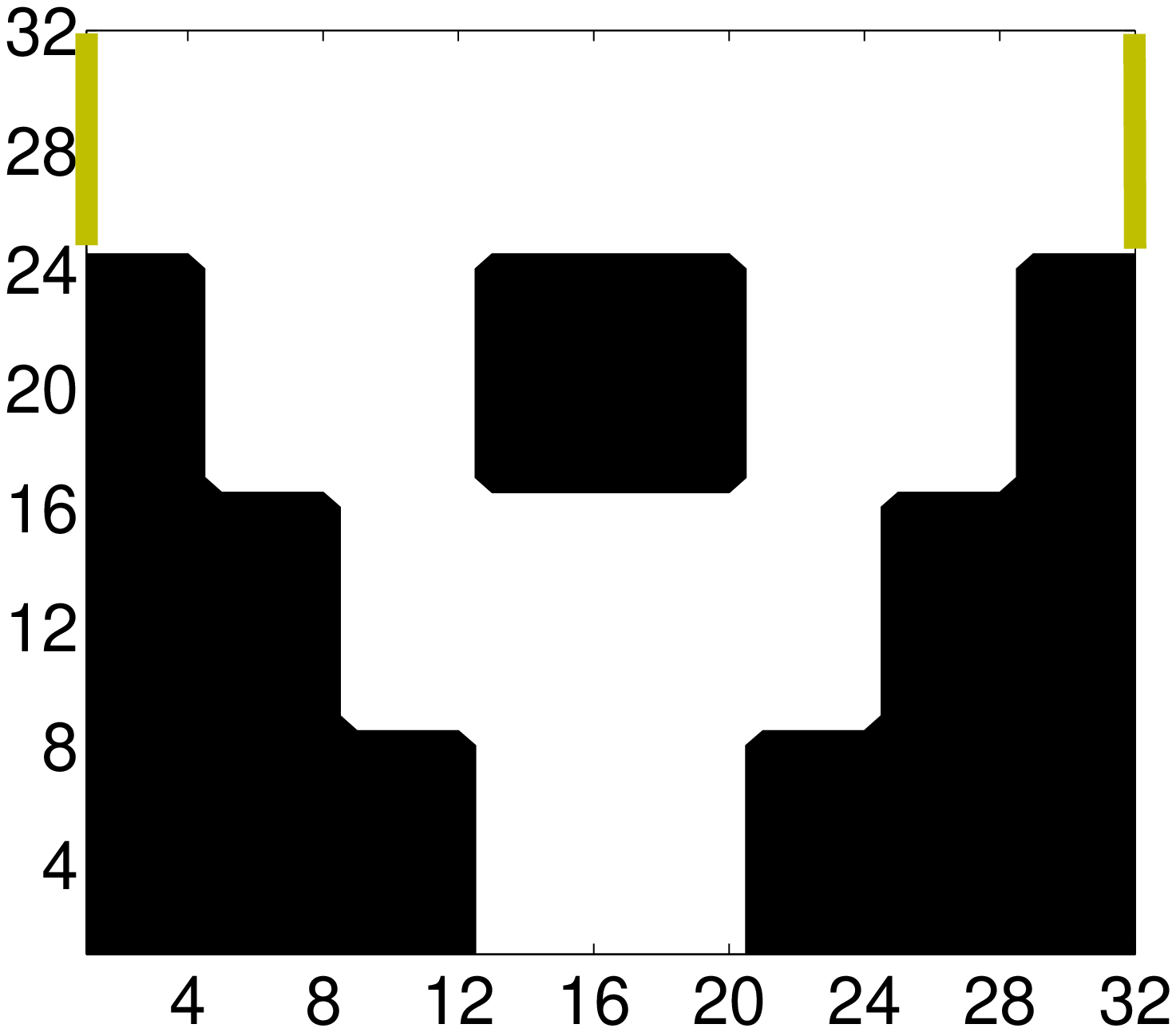}
	%\caption{Case 1}	
\end{minipage}
}    \\  \hspace*{0.01in}
\subfloat[Third generation of SPG.]{
\label{Third generation of SPG.}
\begin{minipage}[h]{0.22\textwidth}
	%\centering
	\includegraphics[width=1.7in]{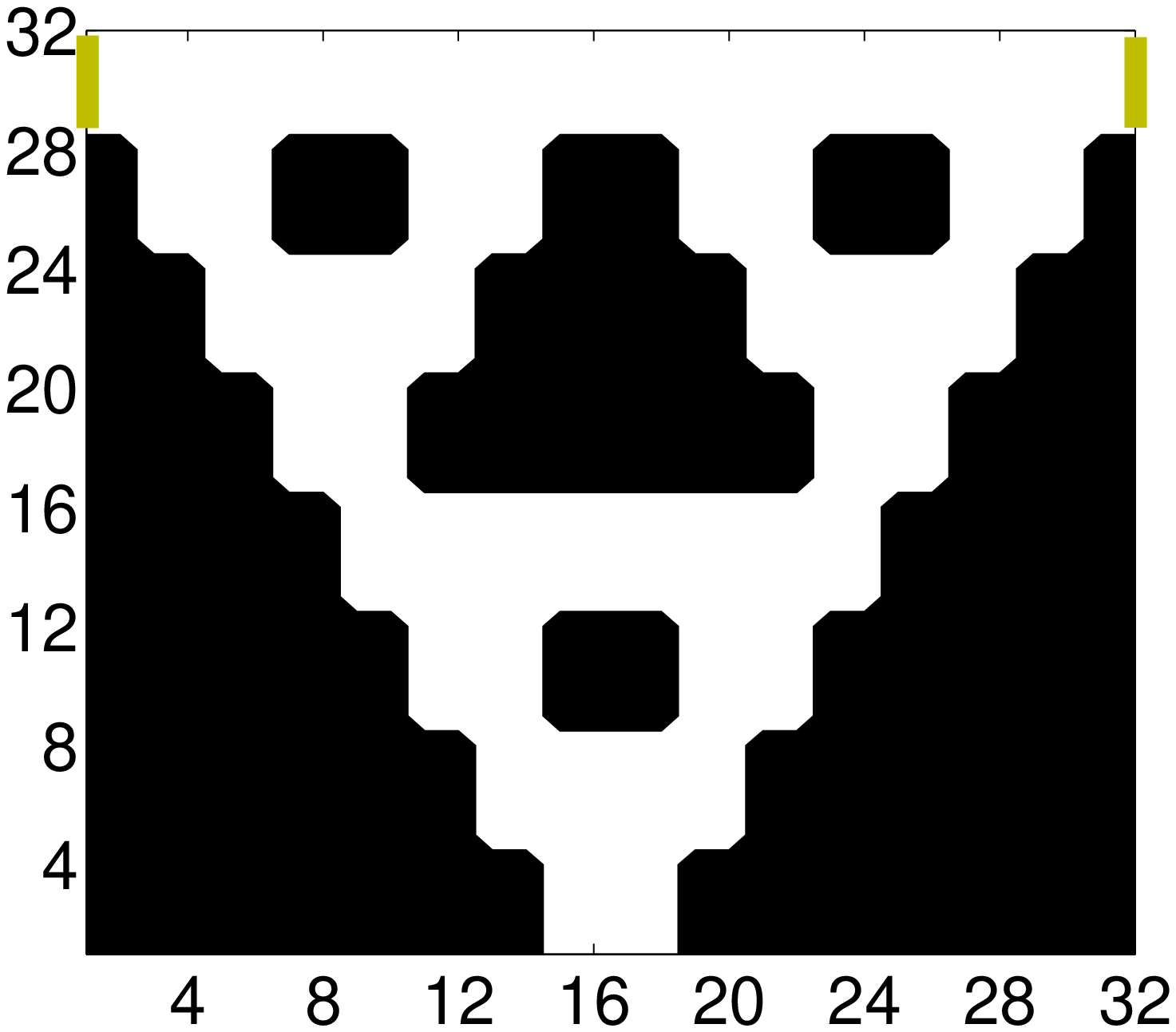}
	%\caption{Case 1}	
\end{minipage}
}
\subfloat[Fourth generation of SPG.]{
\label{Fourth generation of SPG.}
\begin{minipage}[h]{0.22\textwidth}
	%\centering
	\includegraphics[width=1.7in]{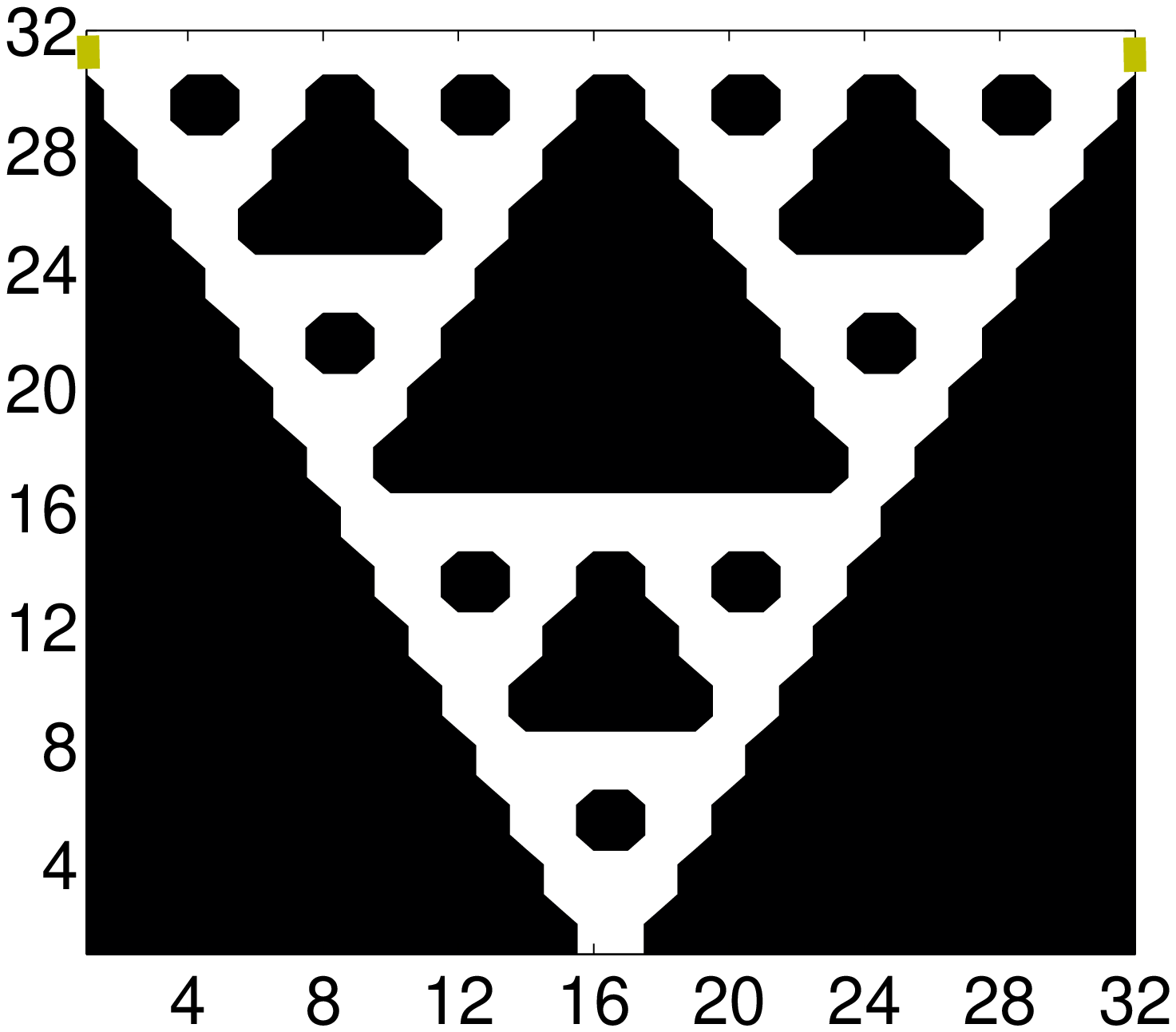}
	%\caption{Case 1}	
\end{minipage}
}
\caption{\raggedright (Color online) Fractal structures of SPGs: an example of planar gasket with the self-similarity of the system. Four SPGs are attached to two semi-infinite metallic leads, left lead and right lead, on the edges (yellow lines). All SPGS are 32 x 32 square lattice and the white regions are the atomic sites. The widths connects with the each two leads are (a)$ 16a $ in first generation of SPG, (b)$ 8a $ in second generation of SPG, (c)$ 4a $ in third generation of SPG and (d)$ 2a $ in fourth generation of SPG.}
\label{fig:fc}
\end{figure}

The Sierpinski gasket is named after the Polish mathematician Sierpinski in 1915 \cite{Sierpinski:1915lr}  and is one of the well-known fractal geometry. Since fractal geometry was established in 1967, there are several ways to obtain Sierpinski gasket (SPG)\cite{PhysRevLett.45.855,Mandelbrot1982,PhysRevB.28.3110,PhysRevLett.52.1033,PhysRevLett.57.649,0295-5075-10-1-013,Sun:2010zr,Sun:2010ly}. Following by previous studies, we choose SPG as our sample and four generations of SPG are shown in Fig. \ref{fig:fc}. The SPG is perfectly self similar, an attribute of many fractal structures. Any triangular portion is an exact replica of the whole structure (Fig. \ref{fig:fc}). The dimension of the SPG is $ log 3 / log 2 = 1.5849 $ and lies dimensionally between a line and a plane.

\subsection{The Landauer Keldysh formalism}
The Landauer Keldysh formalism is also called non-equilibrium Green's function (NEGF) formalism\cite{Keldysh:1965fj} and in principle carries all the physical information of the investigated under a non-equilibrium but approximately steady process\cite{Datta2000253}. It is usually used to simulate current and charge density in atomic-scale quantum mechanical transport when bias applied. Also it can be used to calculate spin density and transmission in different generations of sample. We describe the conductor by an array of numbered lattice sites, $n=1,2, ...,N \times N$. For a $32 \times 32$ sample contains a square lattice of $ 1024 $ sites, as shown in Fig. \ref{fig:fc}. The corresponding Hamiltonian matrix is written in the site basis and each element of matrix $H_{m,n}$ records the relation between sites $m$ and $n$. After we label the sites of lattice, we will show that the interaction between the sites by using the nearest neighbor hopping to describe the interaction, viz, the tight-binding model. With the help of the second quantization, the one-particle Hamiltonian composes of on-site terms and a hopping term, 
\begin{equation}
H=\sum_{n}\varepsilon_{n} c^{\dagger}_{n} c_{n} + \sum_{m,n} c^{\dagger}_{n} t_{m,n} c_{n} \label{eqs:hpure} 
\end{equation}
where $ c^{\dagger}_{n} ( c_{n}  ) $ is the electron creation (annihilation) operator, $\varepsilon_{n}$ is the on-site energy at $ n $-th site, and $ t_{m,n} $ is the hopping matrix. Using the system of the Rashba spin-orbital interaction (RSI)\cite{Bychkov1984} for example, the hopping matrix is
$
t_{m,n} = - t_{0} \textbf{I} - \imath t_{so} (\vec{\sigma} \times \textbf{e}_{n\rightarrow m}) \cdot \textbf{e}_{z}
$
where $ \textbf{I} $ is the $ 2 \times 2 $ identity matrix, $ t_{0} $ is the kinetic hopping strength and $ t_{so} $ is the hopping strength of RSI. Applying the nearest hopping only, one writes
\begin{equation}
H=\sum_{n}\varepsilon_{n} c^{\dagger}_{n} c_{n} + \sum_{m,n} c^{\dagger}_{n} (- t_{0} \textbf{I} - \imath t_{so} (\vec{\sigma} \times \textbf{e}_{n\rightarrow m}) \cdot \textbf{e}_{z}) c_{n} \label{eqs:hso}
\end{equation}
where $ \langle m,n \rangle $ means $m-th$ site interacts with the nearest sites $n$. Further, $ \vert \textbf{r}_{m} - \textbf{r}_{n} \vert = a $ with lattice spacing $a$ is satisfied. Since the Hamiltonain has established, then considering the ideal (free of spin-orbital orbit interactions) leads are connecting with the sample left and right, respectively. Deriving the self-energy from Ref.\cite{Datta:1998fk}, we obtain the self-energy matrix of the left(right) lead, $ \Sigma_{left(right)} $. Next, using both the tight-binding Hamiltonian and lead self-matrices, one can build up the retarded Green function matrix
\begin{equation}
G^{R}(E)=[E\textbf{I}-H-\Sigma_{left}(E)-\Sigma_{right}(E)]^{-1} \label{eqs:rgreen}
\end{equation}
where the $ \textbf{I} $ is $ 2N \times 2N $ identity matrix, $ H $ is the tight-binding Hamiltonian matrix of the conductor, and $ \Sigma_{left(right)}(E) $ is the self-energy matrix of the left(right) lead. Next, the lesser self-energy matrix is given by
\begin{equation}
\Sigma^{<}(E)=-\sum_{p=L, R}[\Sigma_{p}(E-eV_{p}) - \Sigma^{\dagger}_{p}(E-eV_{p})] f_{0}(E-eV_{p})
\end{equation} 
where $ f_{0} $ is the Fermi function and $ eV_{p} $ is the electric bias potential energy applied on lead $ p=\lbrace L(left), R(right)\rbrace $. In the numerical calculation, we set $ eV_{left} = +( e V_{0})/2 $ and $ eV_{right} =-( e V_{0})/2 $. Moreover,we consider the zero temperature simplistically in order to ignore thermal fluctuation. Thus the Fermi function reduces to step function. Finally, we can get the lesser Green function via the kinetic equation 
\begin{equation}
G^{<}(E)=G^{R}(E) \Sigma^{<}(E) G^{A}(E)
\end{equation}
where
$
G^{A}=\lbrace G^{R}\rbrace^{\dagger}
$
Thus we get the lesser Green's function $G^{<}$ which can be used to calculate the local charge density,

%where $eV_{p}$ is the electric potential energy applied on lead $p$.

%\ subsection{Charge Density}
\begin{equation}
e\langle\hat{N}_{n}\rangle=\frac{e}{2\pi i}\int^{\infty}_{-\infty} dE  \; Tr_{s}[G^{<}_{n,n}] \label{eqs:lcd}
\end{equation}
where the trace calculates in spin space and $\hat{N_{n}}$ is the electron number operator at site $n$. Also the local spin density will be
%\ subsection{Spin Density}
\begin{equation}
\langle\vec{S_{n}}\rangle=\frac{\hbar}{2}\frac{1}{2\pi i}\int^{\infty}_{-\infty} dE \; Tr_{s}[\vec{\sigma}G^{<}_{n,n}] \label{eqs:lsd}
\end{equation}
here $\vec{S_{n}}=(S_x, S_y, S_z)$ is spin operator at site $n$ of the conductor. We assume our devices lay on $ x $ and $ y $ directions. Thus, the current goes along in x-direction. The study of the latter, the local spin density, can be categorized into the out-plane $ \left( \left\langle S_{z} \right\rangle = \left\langle \textbf{S}_{\perp} \right\rangle  \right) $ and in-plane $ \left(  \left\langle S_{x},S_{y} \right\rangle = \left\langle \textbf{S}_{\parallel}\right\rangle  \right) $ components. And we set the hopping energy $ t_{0}=1  $, the mass of electron $ m_{e}=1  $, the charge of electron $ q=1 $, and the Plank's constant divided by $ 2\pi $, $\frac{\hbar}{2\pi}=1 $, as the units. We also can calculate transmission,

%\ subsection{Transmission}

\begin{equation}
T=Tr_{s}[\Gamma_{left} G^{R} \Gamma_{right} G^{A}]
\end{equation}
where $ \Gamma_{left(right)} = \imath [ \Sigma_{left(right)}-\Sigma_{left(right)}^{\dagger}]$ describes the coupling of the conductor to the two leads.

\section{Numerical Results and Discussions}

\subsection{First Generation of SPG}
\label{sec:fc1}
\begin{figure}[!h]
\centering
\subfloat[The local charge density]{
\label{fig:ql:c1}
\begin{minipage}[t]{0.22\textwidth}
	\centering
	\includegraphics[width=1.7in]{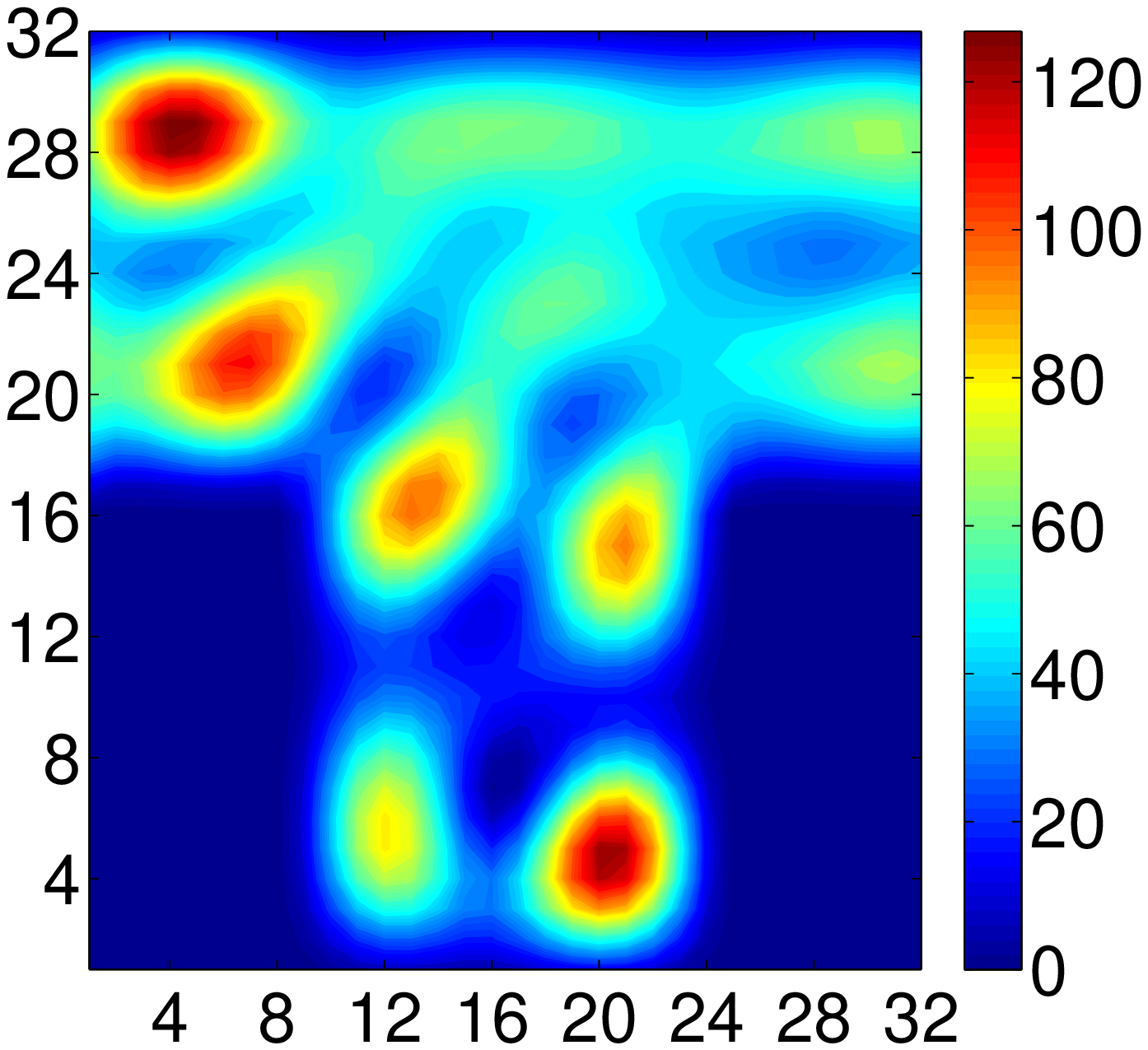}	
	%\caption{Case 2}
\end{minipage}
}
\subfloat[$ S_{z} $]{
\label{fig:szl:c1}
\begin{minipage}[t]{0.22\textwidth}
	%\centering
	\includegraphics[width=1.7in]{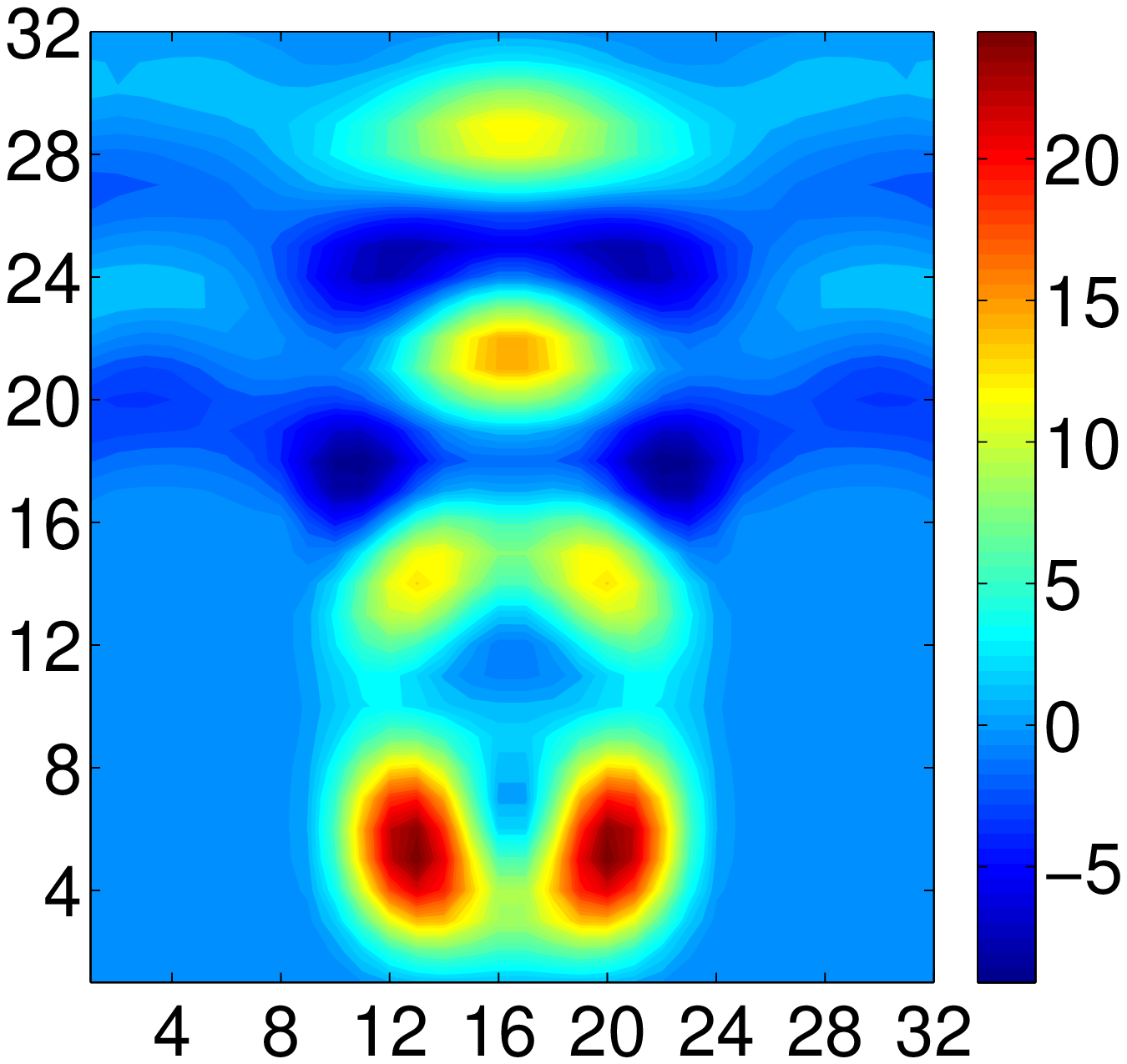}
	%\caption{Case 3}	
\end{minipage}
}
\caption{\raggedright (Color online) Local spin density and charge density under a low bias in first generation. In Fig. \ref{fig:szl:c1}, there is a linear wire interacting with a quantum dot. }
\label{fig:sdql:c1}
%\begin{figure*}[!t]
%\centerline{\subfloat[Case I]\includegraphics[width=2.5in]{subfigcase1}%
%\hfil
%\subfloat[Case II]{\includegraphics[width=2.5in]{subfigcase2}%
%\label{fig_second_case}}}
%\caption{Simulation results}
%\label{fig_sim}
%\end{figure*}
%
\end{figure}

\begin{figure}[!h]
%\centering
\subfloat[The local charge density]{
\label{fig:qh:c1}
\begin{minipage}[h]{0.22\textwidth}
	%\centering
	\includegraphics[width=1.7in]{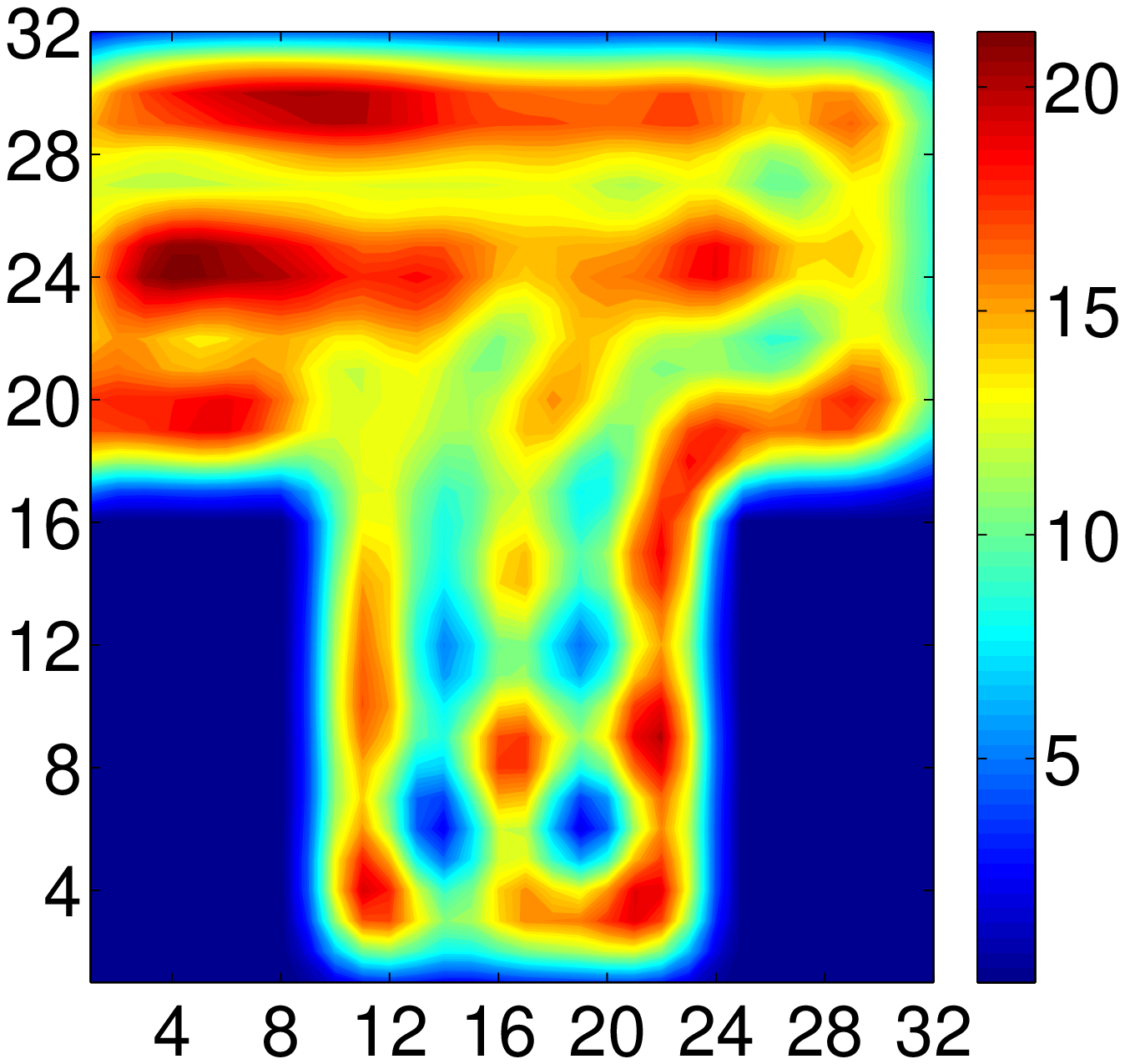}	
	%\caption{Case 2}
\end{minipage}
}
\subfloat[$ S_{z} $]{
\label{fig:szh:c1}
\begin{minipage}[h]{0.22\textwidth}
	%\centering
	\includegraphics[width=1.7in]{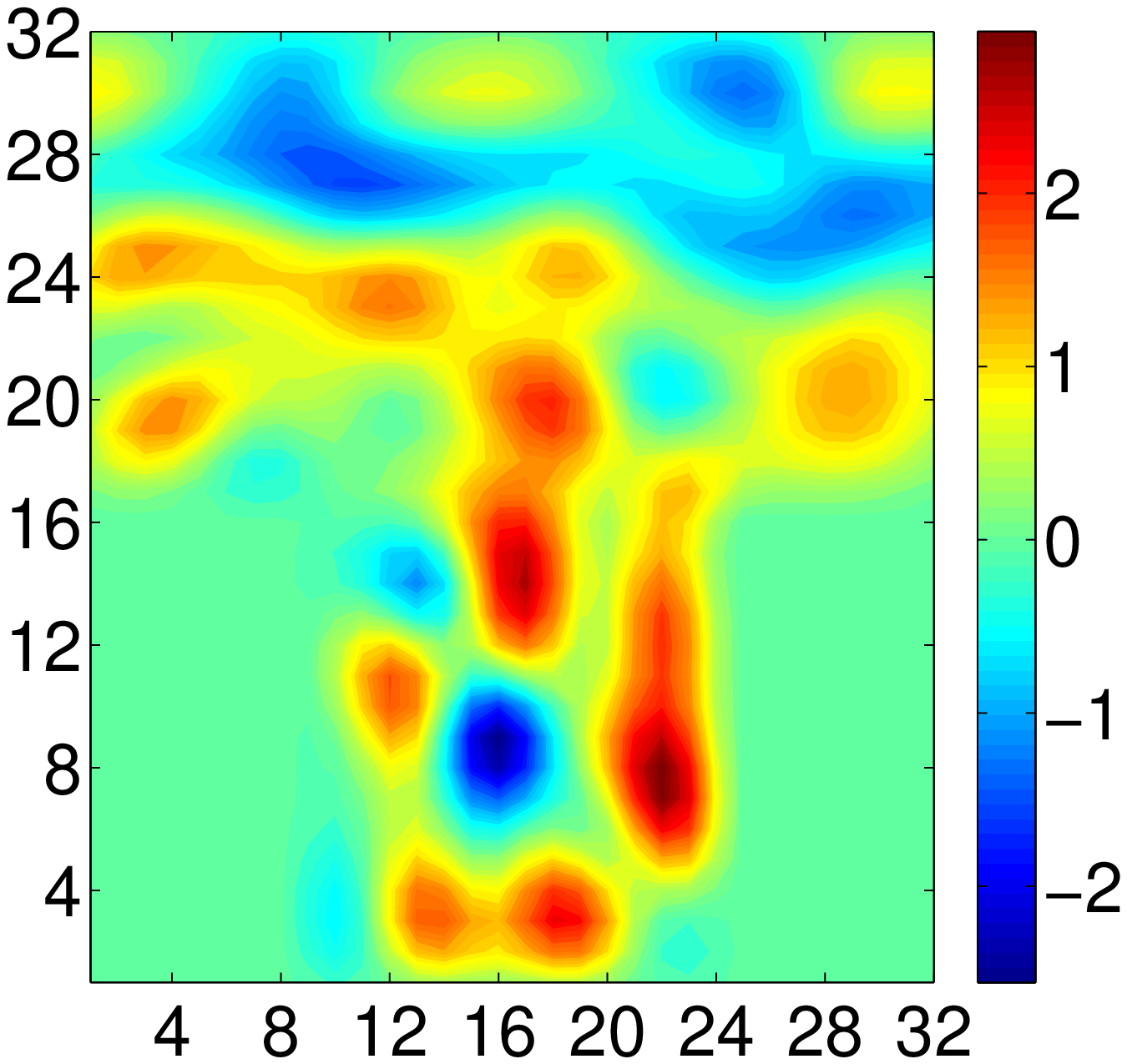}
	%\caption{Case 3}	
\end{minipage}
}
\caption{\raggedright (Color online) Local spin density and charge density under a strong bias in first generation.}
\label{fig:sdqh:c1}
\end{figure}

The first generation of SPG is setting in the band bottom $ E_{b}=-4t_{0} $ and the Fermi-level $ E_{f}=-3.8t_{0} $. Here, the black regions of the sample are empty sites and the white regions are atomic sites for transporting charges and spins, as shown in Fig. \ref{fig:sdql:c1}. The width connects with the each two leads is $ 16a $. \\
From Fig. \ref{fig:sdql:c1}, under low bias situation, the charge transport separated into two channels and the local charge accumulate in the bottom channel. The first generation of SPG look like a combination of linear conducting line coupled with a localized quantum dot(QD). A QD is a part of material confined in three dimensions\cite{1981JETPL..34..345E,2000AnRMS..30..545M}. Fig. \ref{fig:szl:c1} clearly indicates the up-spin electrons mainly localized at QD-like region and the down-spin electrons is the conducting carrier for first generation of SPG. The separation of up-spin electrons and down-spin electrons are resulted from the coupling of RSI. Therefore, first generation of SPG behave as good spin filter under RSI.\\
When raising into high bias, electrons with high energy injected from the lead and transport through the device. Now high energy electrons can go from the linear region into the QD-like region and then back to the linear region again (Fig. \ref{fig:qh:c1}). Therefore, charge did not localized in the QD-like region as in Fig. \ref{fig:ql:c1}, and conducting path penetrating into the QD-like region, i.e., the bottom of the device. Comparing with the low bias situation, both the down-spin and up-spin electrons can form conducting channels. While within the QD-like region, there are still mainly up-spin electrons and only a few down-spin electrons trapped in the center of the bottom of the device at high bias (Fig. \ref{fig:szh:c1})\\
In first generation of SPG, quantum dot-like region occurs at low bias. Beside the bottom square shape of the first generation of SPG is QD-like and spin-polarized electrons driven by the RSI will let the up-spin electrons trapped within the QD-like regions. Therefore, a spin-dependent potential also form within the QD region and thus enhance the confinemnt of up-spin electrons. Therefore, first generation of SPG behaves as good spin filter with RSI.
% 補上 the condition of quantum dot

\subsection{Second Generation of SPG}
\label{sec:fc2}
\begin{figure}[!h]
%\centering
\subfloat[The local charge density]{
\label{fig:ql:c2}
\begin{minipage}[h]{0.22\textwidth}
	%\centering
	\includegraphics[width=1.7in]{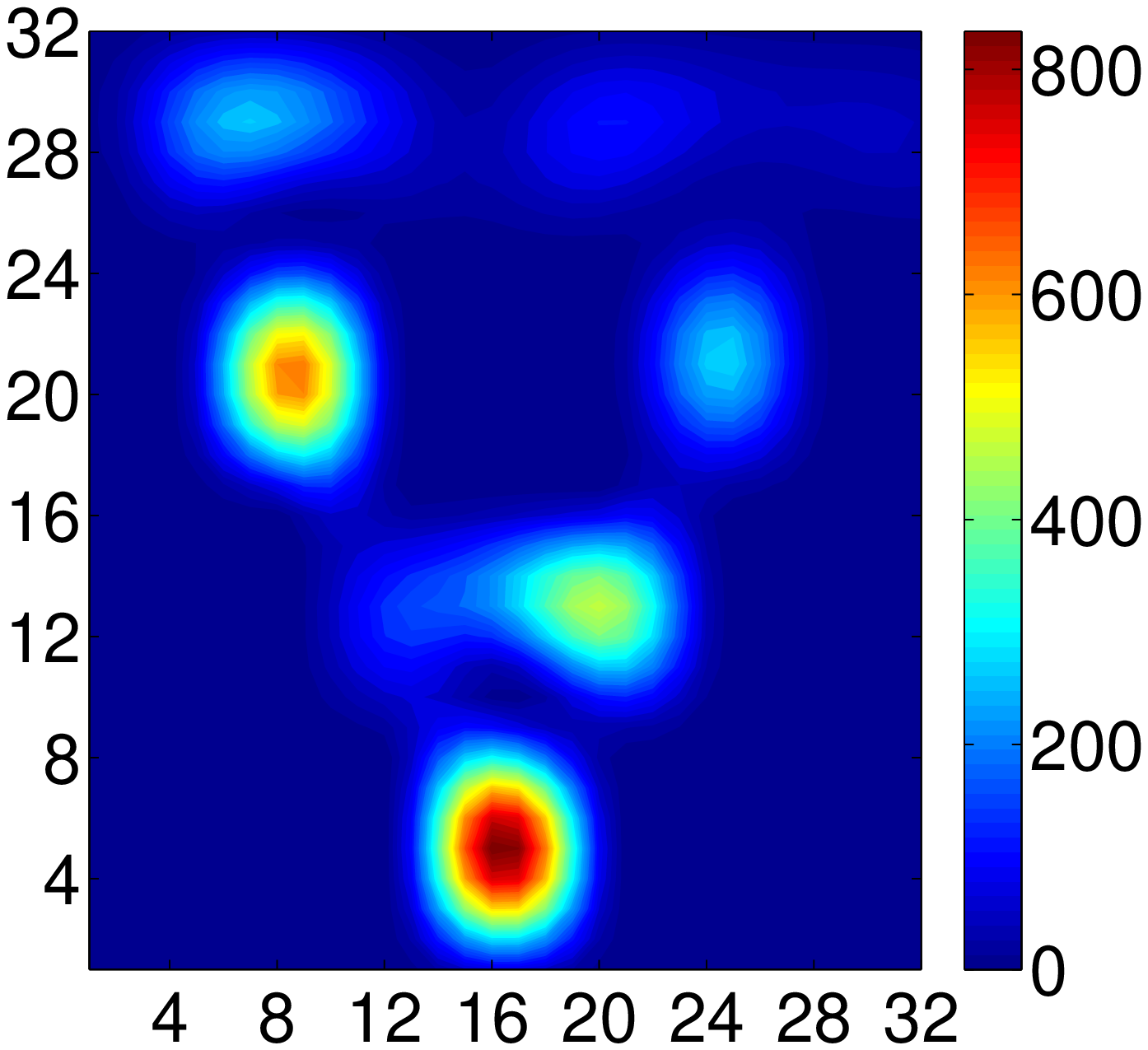}	
	%\caption{Case 2}
\end{minipage}
} 
\subfloat[$ S_{z} $]{
\label{fig:szl:c2}
\begin{minipage}[h]{0.22\textwidth}
	%\centering
	\includegraphics[width=1.7in]{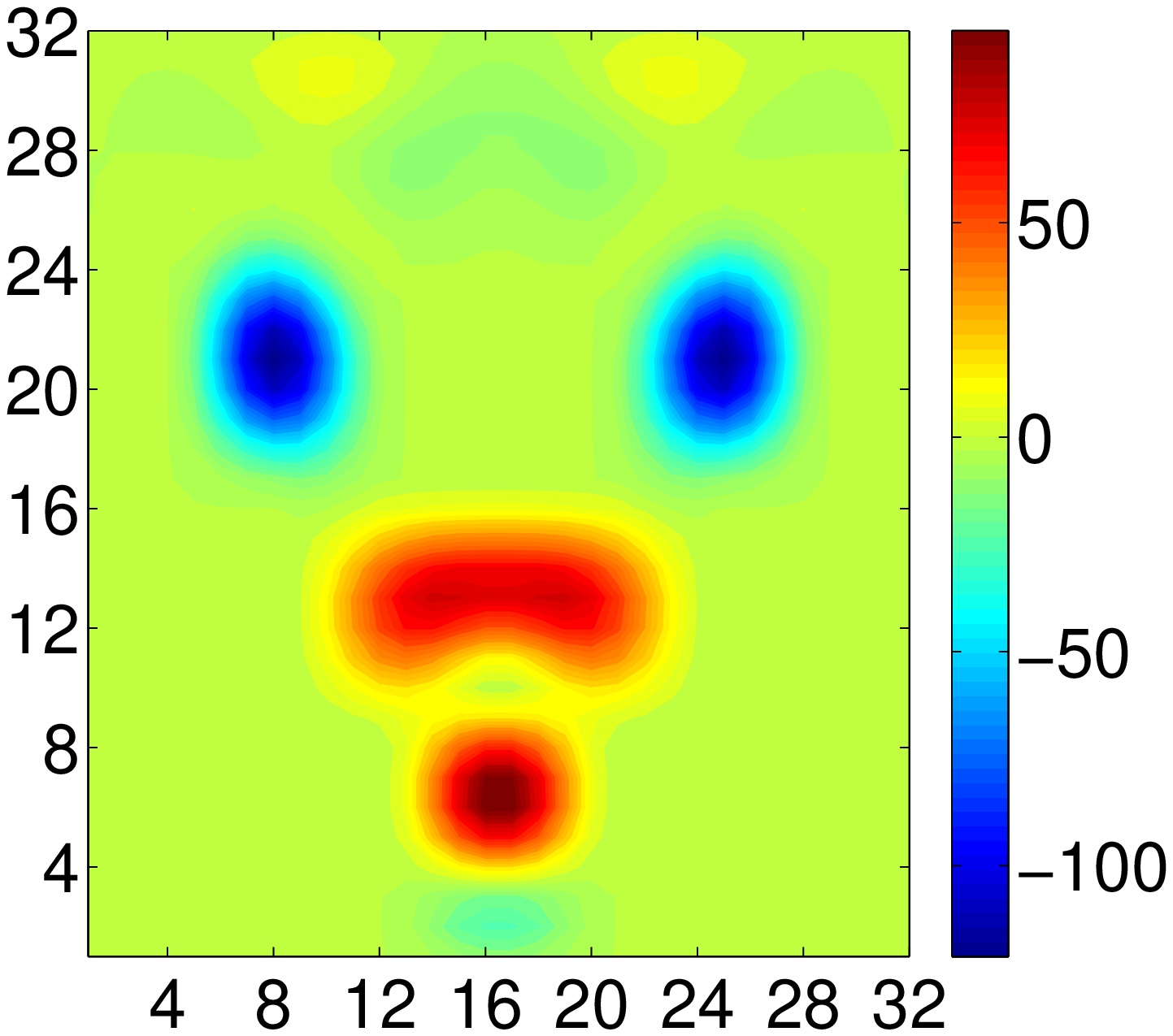}
	%\caption{Case 3}	
\end{minipage}
}
\caption{\raggedright (Color online) Local spin density and charge density under a low bias in second generation.}
\label{fig:sdql:c2}
\end{figure}

\begin{figure}[!h]
%\centering 
\subfloat[The local charge density]{
\label{fig:qh:c2}
\begin{minipage}[h]{0.22\textwidth}
	%\centering
	\includegraphics[width=1.7in]{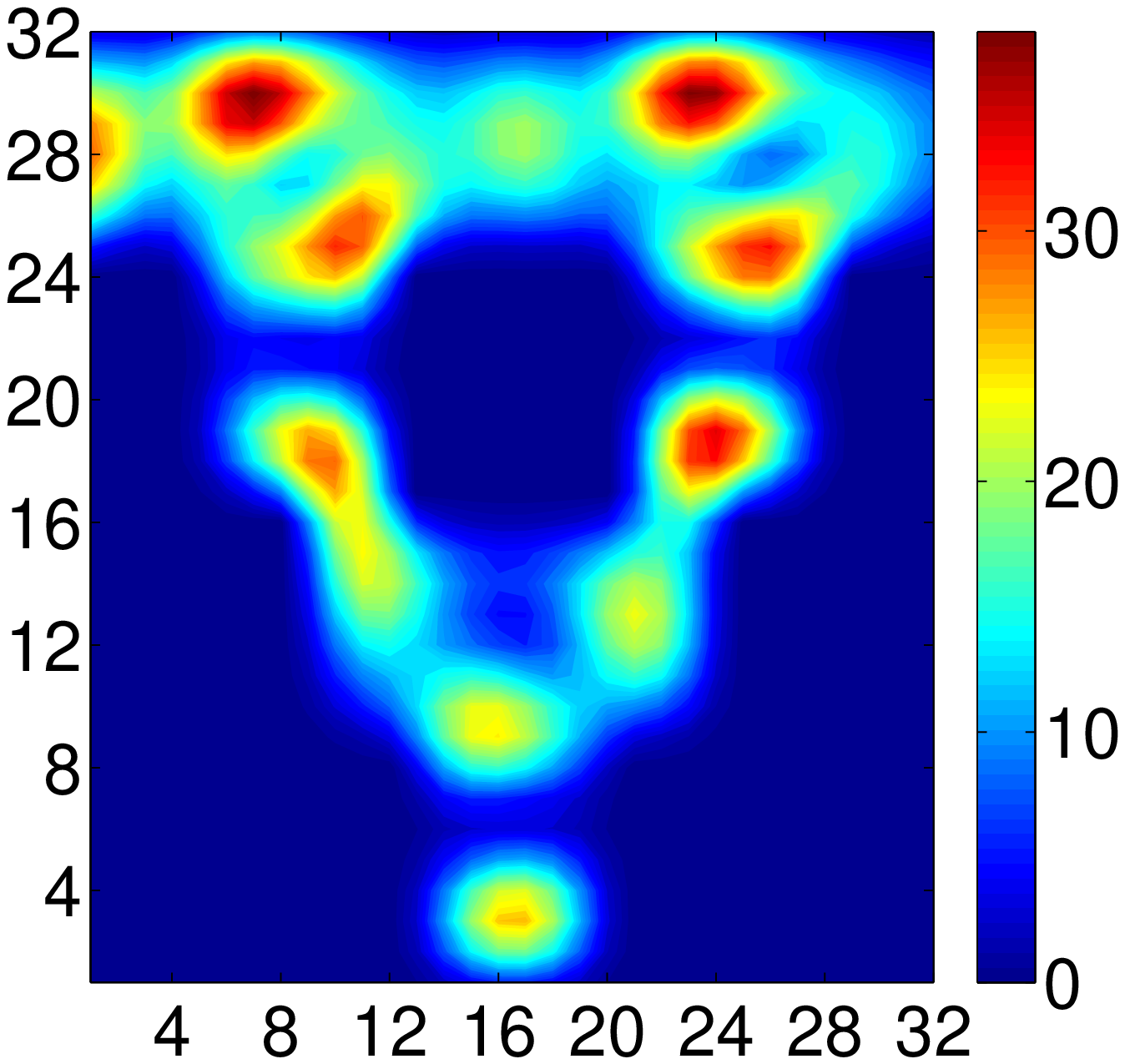}	
	%\caption{Case 2}
\end{minipage}
} 
\subfloat[$ S_{z} $]{
\label{fig:szh:c2}
\begin{minipage}[h]{0.22\textwidth}
	%\centering
	\includegraphics[width=1.7in]{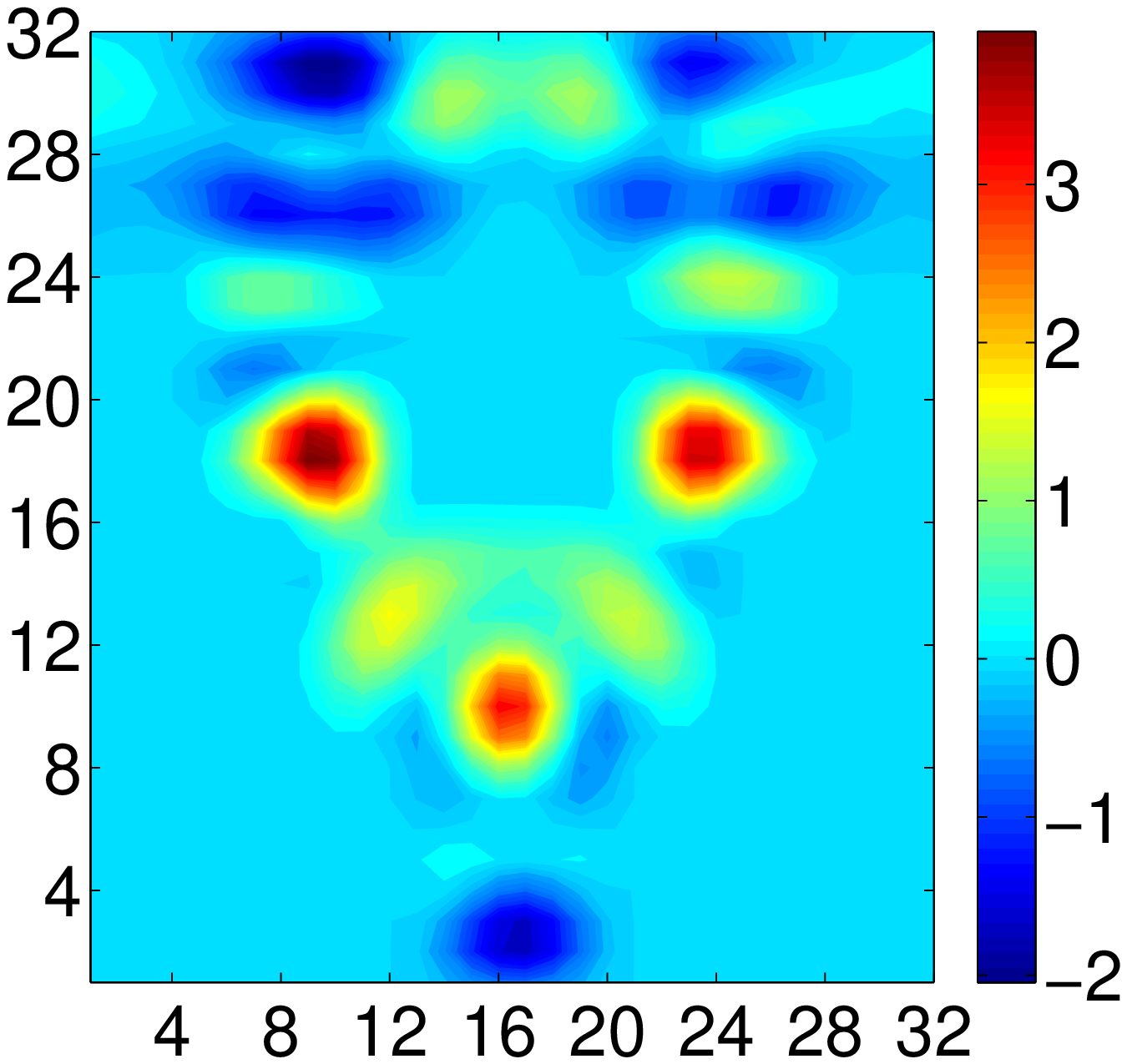}
	%\caption{Case 3}	
\end{minipage}
}
\caption{\raggedright (Color online) Local spin density and charge density under a strong bias in second generation.}
\label{fig:sdqh:c2}
\end{figure}

The second generation of SPG (Fig. \ref{Second generation of SPG.}) is composed of three self-similar building blocks of first generation (Fig. \ref{First generation of SPG.}). There are empty sites (black regions) and atomic sites (white regions) in the device. The conditions are setting in the band bottom $ E_{b}=-4t_{0} $ and the Fermi-level $ E_{f}=-3.8t_0 $, as the same as first generation. However, the connecting width with two leads is different and is $ 8a $ instead of $ 16a $.\\
The second generation of SPG geometrically look like three first generation of SPG connects in series (Fig. \ref{First generation of SPG.} and Fig. \ref{Second generation of SPG.}). However for low bias situation, the electrons are mainly accumulated in the bottom of the device in second generation of SPG (Fig. \ref{fig:ql:c2}). There is almost no clear conducting channel can be observed with the low bias. Additionally, comparing with the first generation of SPG, the second of SPG is topological different from the first generation of SPG with a hole at the center of fractal structure. As reported in Ref.\cite{chen:07E908}, a point-defect induced symmetry breaking significantly changes the conduction channel. The charge density seems to be in favor to pass through the bottom of the device, instead of through the upper line. The width of the top region of the device can change the conducting path of device\cite{chen:07E908,Chen:qy}. The local charge density indicated the charge accumulates in corners of the building blocks in Fig. \ref{fig:ql:c2}. Local spin density also indicated that the spin actually associated with charge and accumulated at the same sites but with alternating polarization of the spin (Fig. \ref{fig:szl:c2}). Nevertheless, both local charge density and local spin density show that the second generation of SPG is much localized than the first generation of SPG. A possible reason is from the interference of spin-polarized electrons through the channels\cite{PhysRevB.77.045324} around the central holes and thus the conducting paths almost disappear in low bias. It should be noted that there is also no clearly QD-like behavior in second generation. Focusing on $ S_z $, the $ 2^{nd} $ generation of SPG can be interpreted as the superposition of three building blocks of $ 1^{st} $ generation of SPG (Fig. \ref{fig:szl:c1} and Fig. \ref{fig:ql:c1}) but with one bottom down-spin block and two up-spin blocks atop (Fig. \ref{fig:szl:c2}).\\
As raising higher bias, QD-like behavior appear at the bottom building block of the device (Fig. \ref{fig:szh:c2}) but similar QD-like behavior does not occur in the top two building blocks. The condition of confinement is highly required without electron leaking. Here a higher bias did not provide enough energy for the electrons hopping into the bottom building block and also jumping out again. Therefore, a QD-like behavior only observed in bottom building block. Even though the geometrical similarity within SPG, the electrodes only connects to the left and right ends of the upper building blocks and thus the similarity of three building blocks was removed by the bias. Therefore, the top two blocks become conducting channels and the bottom one acts as QD-like region. We also noted that the Fermi wavelength and the unit length of building blocks are also critical for the conducting channels, As the conducting distance between two leads equals to the multiple of Fermi wavelength, several oscillating periods of local spin density and local charge density can be form in SPG\cite{chen:07B721}.

\subsection{Third Generation of SPG}
\label{sec:fc3}
\begin{figure}[!h]
%\centering 
\subfloat[The local charge density]{
\label{fig:ql:c3}
\begin{minipage}[h]{0.222\textwidth}
	%\centering
	\includegraphics[width=1.7in]{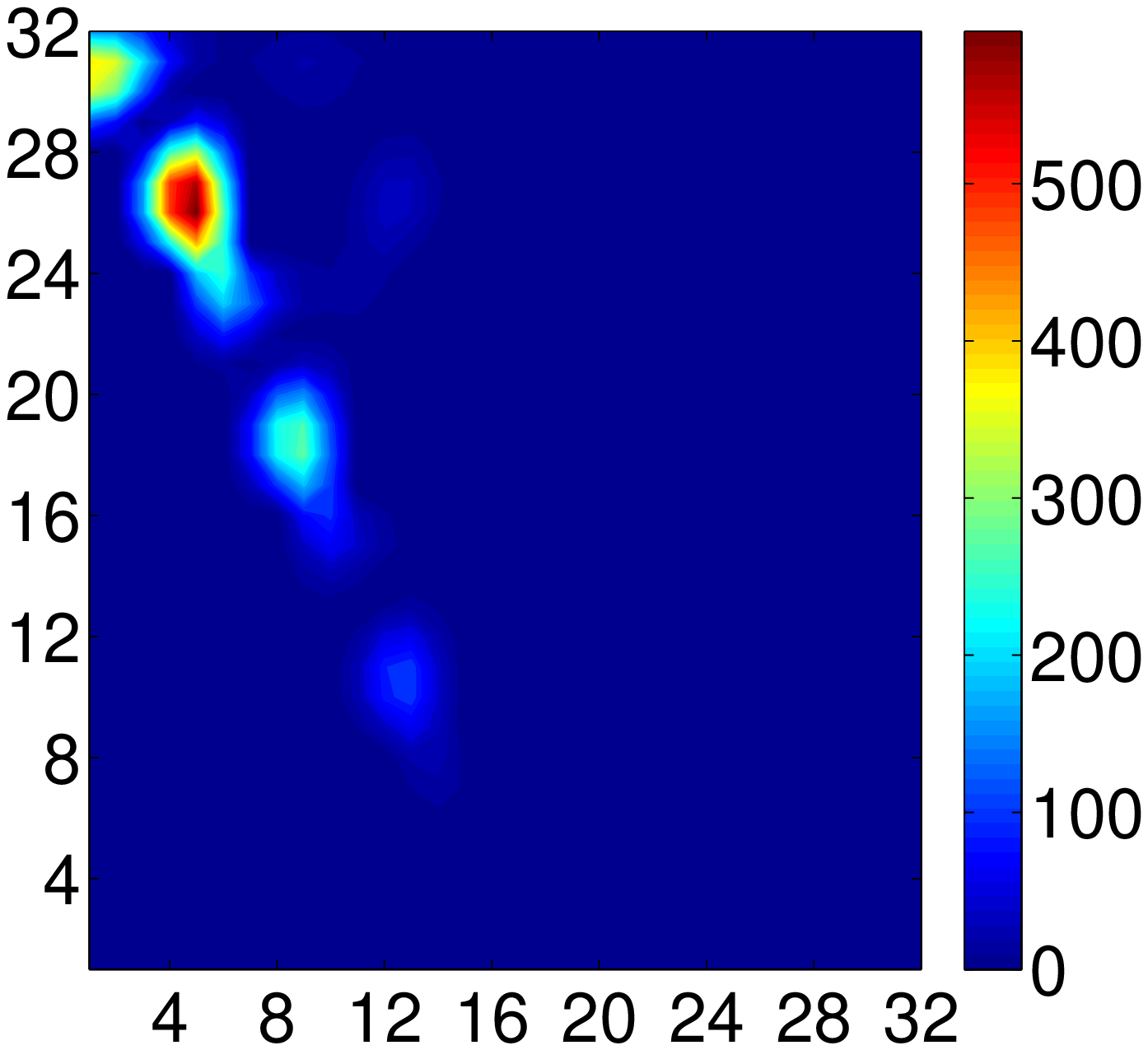}	
	%\caption{Case 2}
\end{minipage}
} 
\subfloat[$ S_{z} $]{
\label{fig:szl:c3}
\begin{minipage}[h]{0.222\textwidth}
	%\centering
	\includegraphics[width=1.7in]{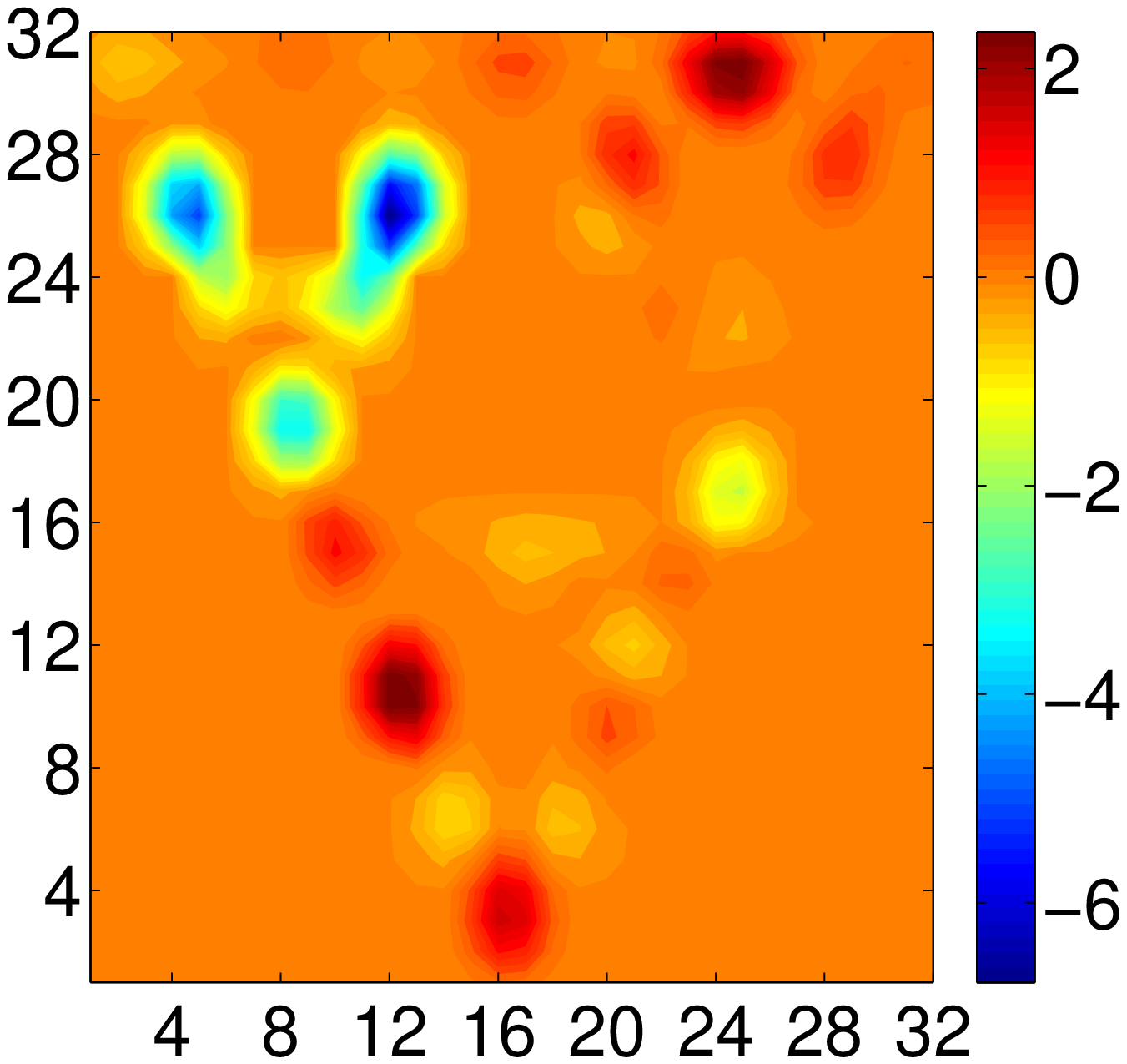}
	%\caption{Case 3}	
\end{minipage}
}
\caption{\raggedright (Color online) Local spin density and charge density under a low bias and $ E_{f}=-3.69 $ in third generation.}
\label{fig:sdql:c3}
\end{figure}

\begin{figure}[!h]
%\centering 
\subfloat[The local charge density]{
\label{fig:qh:c3}
\begin{minipage}[h]{0.22\textwidth}
	%\centering
	\includegraphics[width=1.7in]{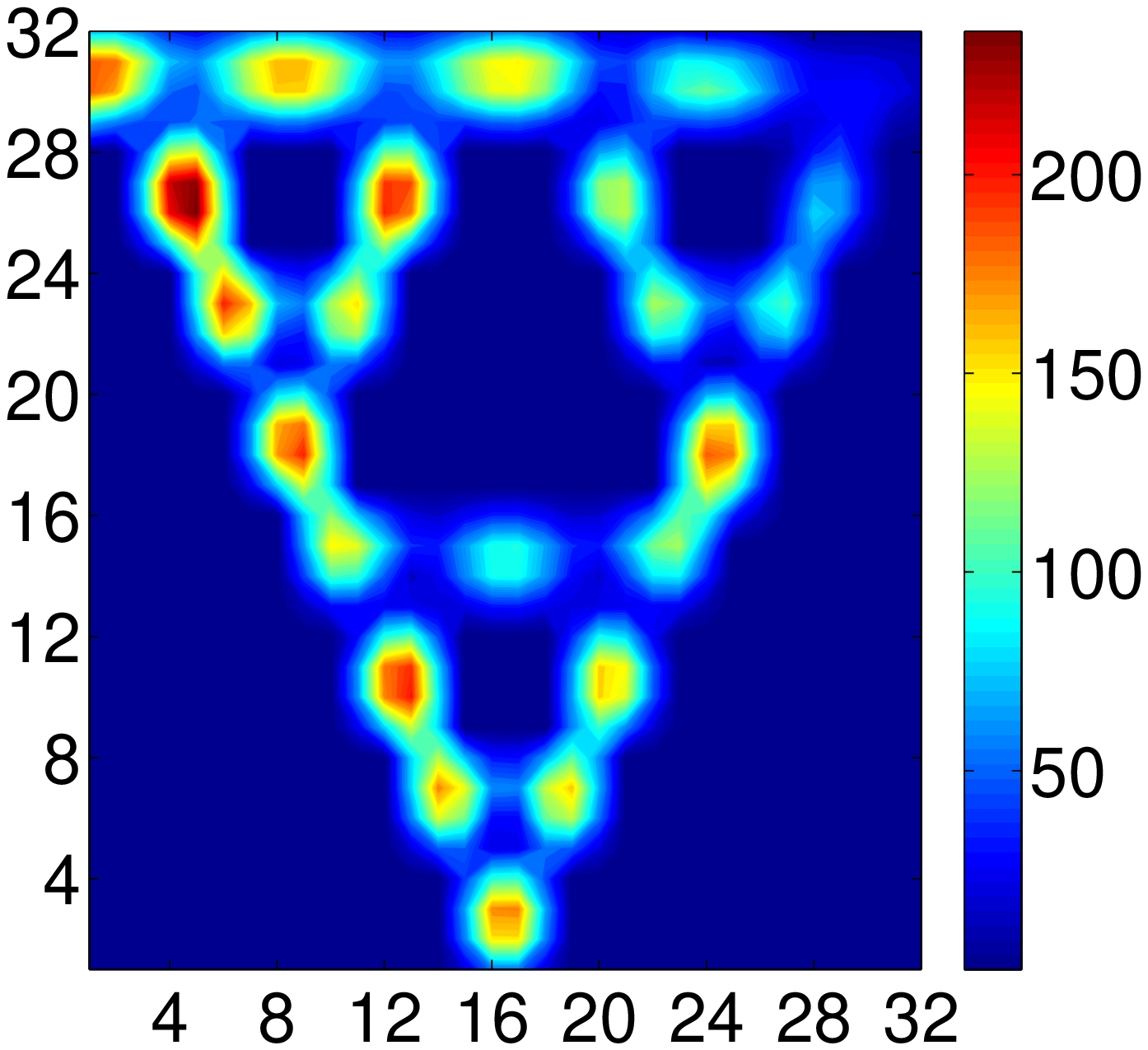}	
	%\caption{Case 2}
\end{minipage}
} 
\subfloat[$ S_{z} $]{
\label{fig:szh:c3}
\begin{minipage}[h]{0.22\textwidth}
	%\centering
	\includegraphics[width=1.7in]{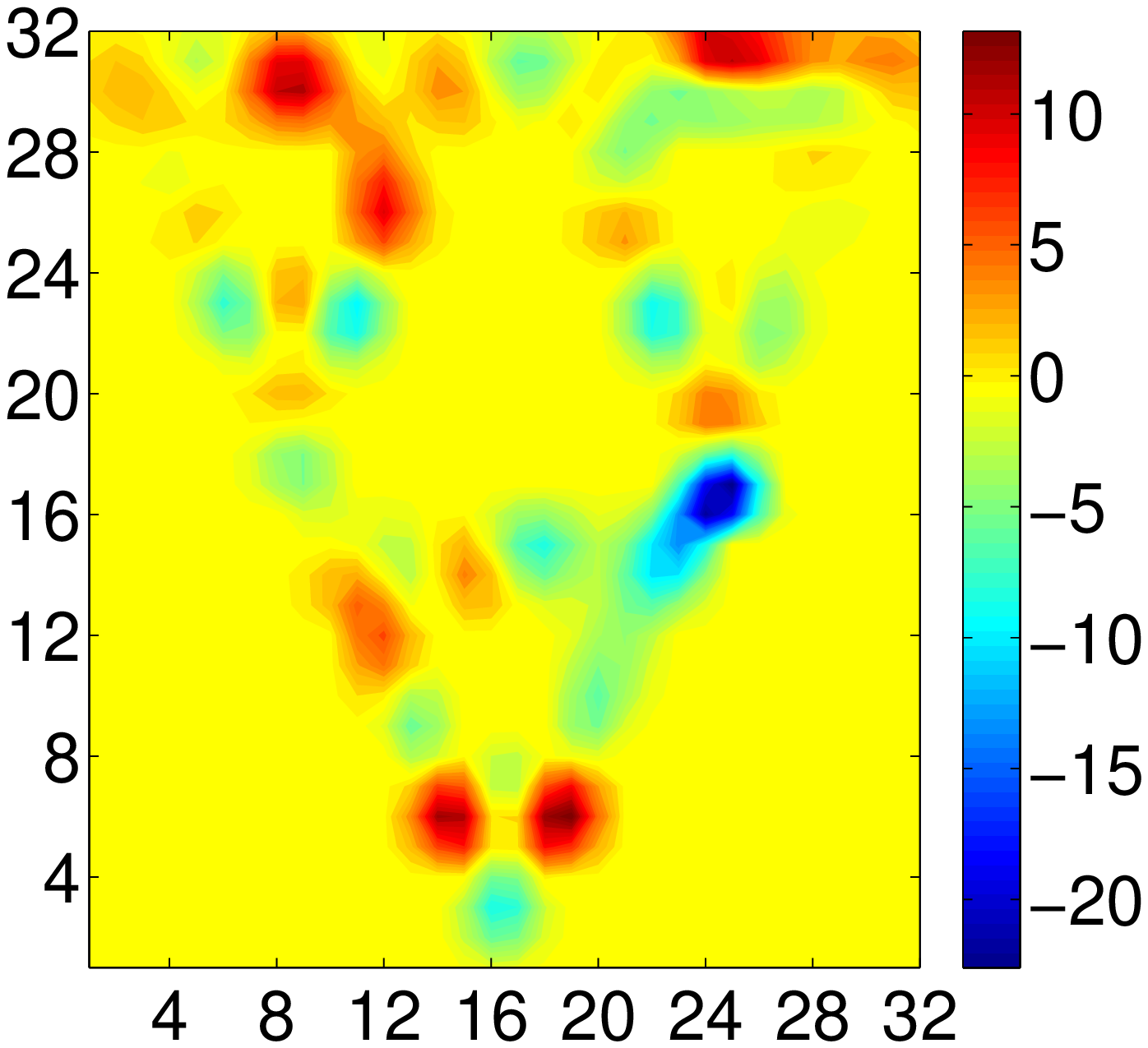}
	%\caption{Case 3}	
\end{minipage}
}
\caption{\raggedright (Color online) Local spin density and charge density under a strong bias  and $ E_{f}=-3.39 $ in third generation.}
\label{fig:sdqh:c3}
\end{figure}

The $ 3^{rd} $ generation is setting in the band bottom $ E_{b}=-4t_0 $ and the Fermi-level $ E_{f}=-3.39t_0 $. Here the sites of the lattice are reduced much more (Fig. \ref{Third generation of SPG.}) and the width connecting with two leads become $ 4a $, so that the Fermi-level increases. In $ 3^{rd} $ generation, the central defect enlarges and there are also other smaller defects. Therefore there are many vortexes around the building blocks. Spin and charge are accumulated at the corners of building blocks. Even though the geometrical self-similarity of the $ 3^{rd} $ generation of SPG, due to the complicated paths, the transports of self-similarity of the $ 1^{st} $ and $ 2^{nd} $ generation of SPG are hardly to see even under high bias in Fig. \ref{fig:szh:c3}. Under low bias, the charge density appears only in the left of the device and it indicates that the bias is weak enough to drive the charge to the other end of SPG for too many empty sites (Fig. \ref{fig:ql:c3}). For a high bias, local charge density is around the empty sites, however, it does not directly go along the narrow line to the other end (Fig. \ref{fig:qh:c3}). It also noted that the local spin density shows only the up-spin electrons were selected to transport within this downward triangular SPG (Fig. \ref{fig:szh:c3}).

\subsection{Fourth Generation of SPG}
\label{sec:fc4}
\begin{figure}[!h]
%\centering 
\subfloat[Local charge density]{
\label{fig:ql:c4}
\begin{minipage}[h]{0.22\textwidth}
	%\centering
	\includegraphics[width=1.7in]{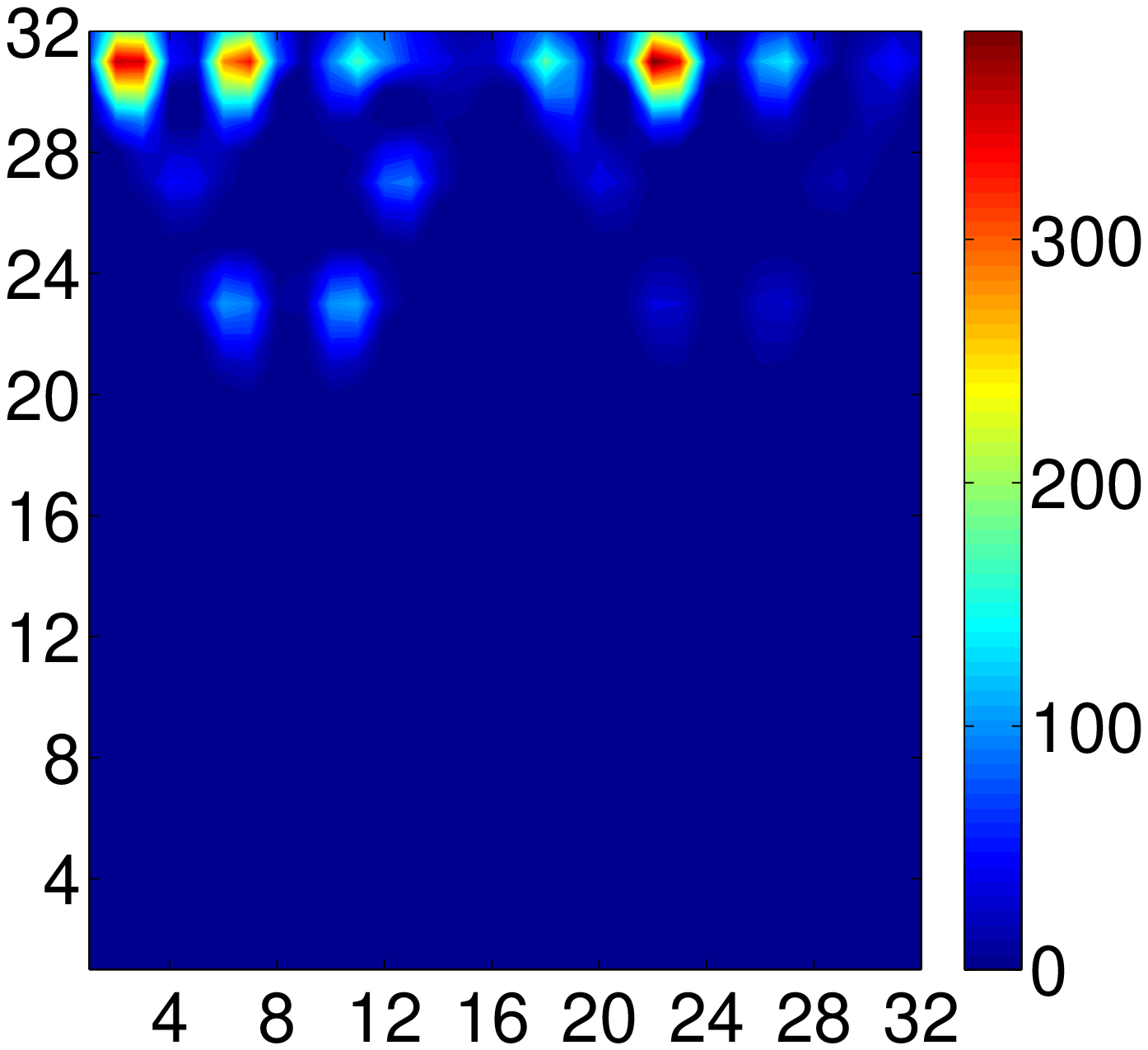}	
	%\caption{Case 2}
\end{minipage}
} 
\subfloat[$ S_{z} $]{
\label{fig:szl:c4}
\begin{minipage}[h]{0.22\textwidth}
	%\centering
	\includegraphics[width=1.7in]{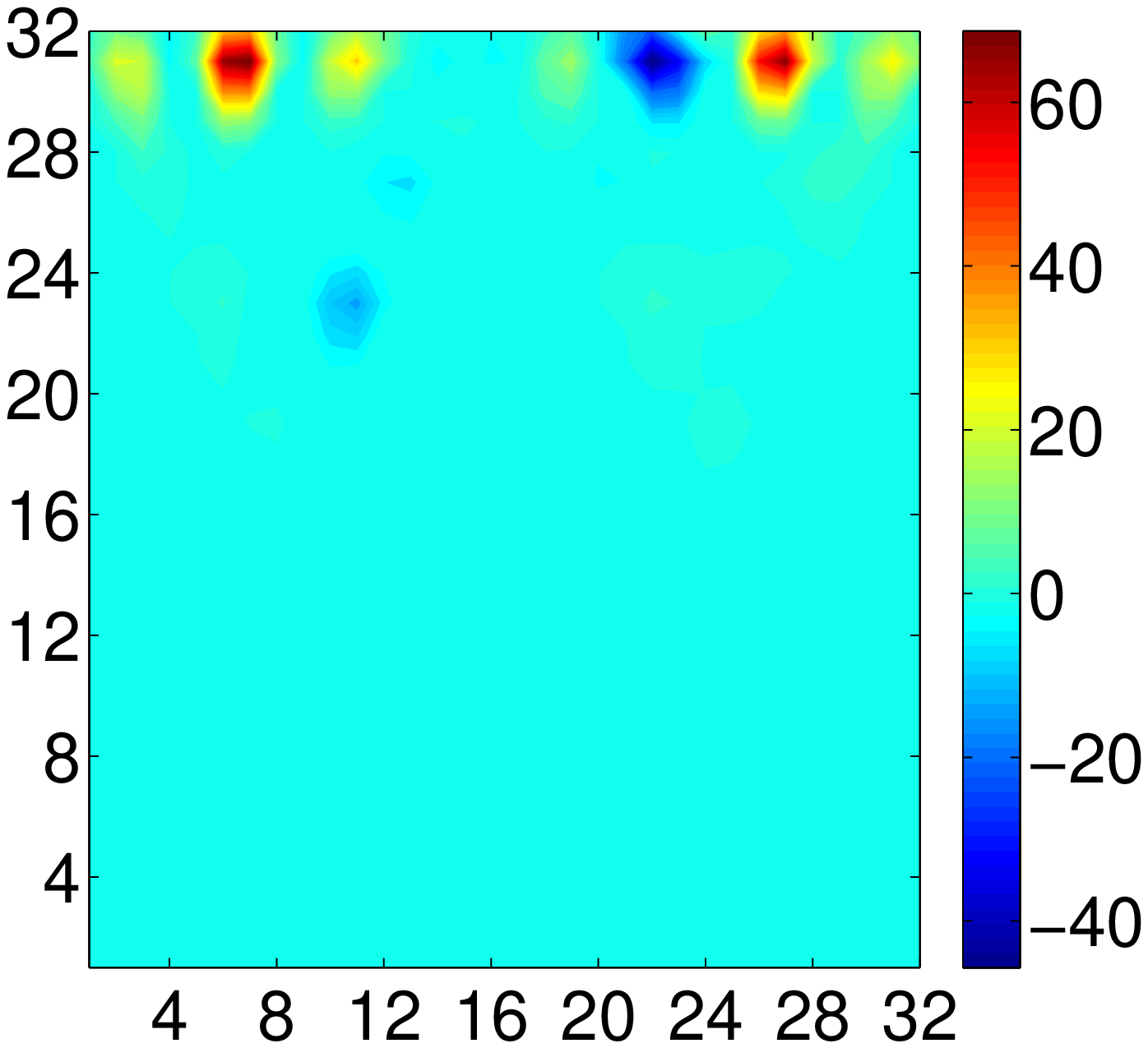}
	%\caption{Case 3}	
\end{minipage}
}
\caption{\raggedright (Color online) Local spin density and charge density under a low bias and $ E_{f}=-2.9 $ in fourth generation.}
\label{fig:sdql:c4}
\end{figure}

\begin{figure}[!h]
%\centering 
\subfloat[Local charge density]{
\label{fig:qh:c4}
\begin{minipage}[h]{0.22\textwidth}
	%\centering
	\includegraphics[width=1.7in]{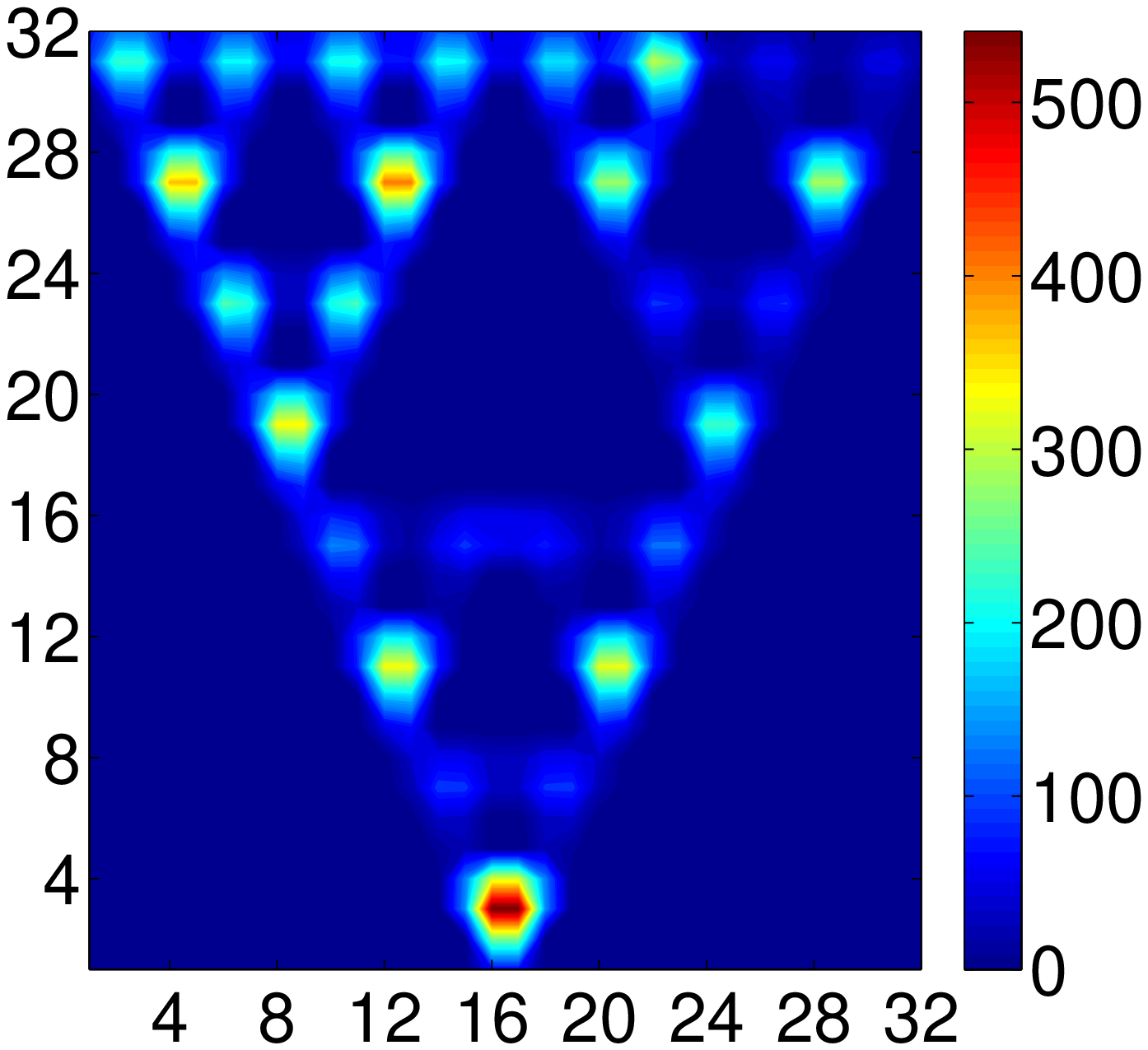}	
	%\caption{Case 2}
\end{minipage}
} 
\subfloat[$ S_{z} $]{
\label{fig:szh:c4}
\begin{minipage}[h]{0.22\textwidth}
	%\centering
	\includegraphics[width=1.7in]{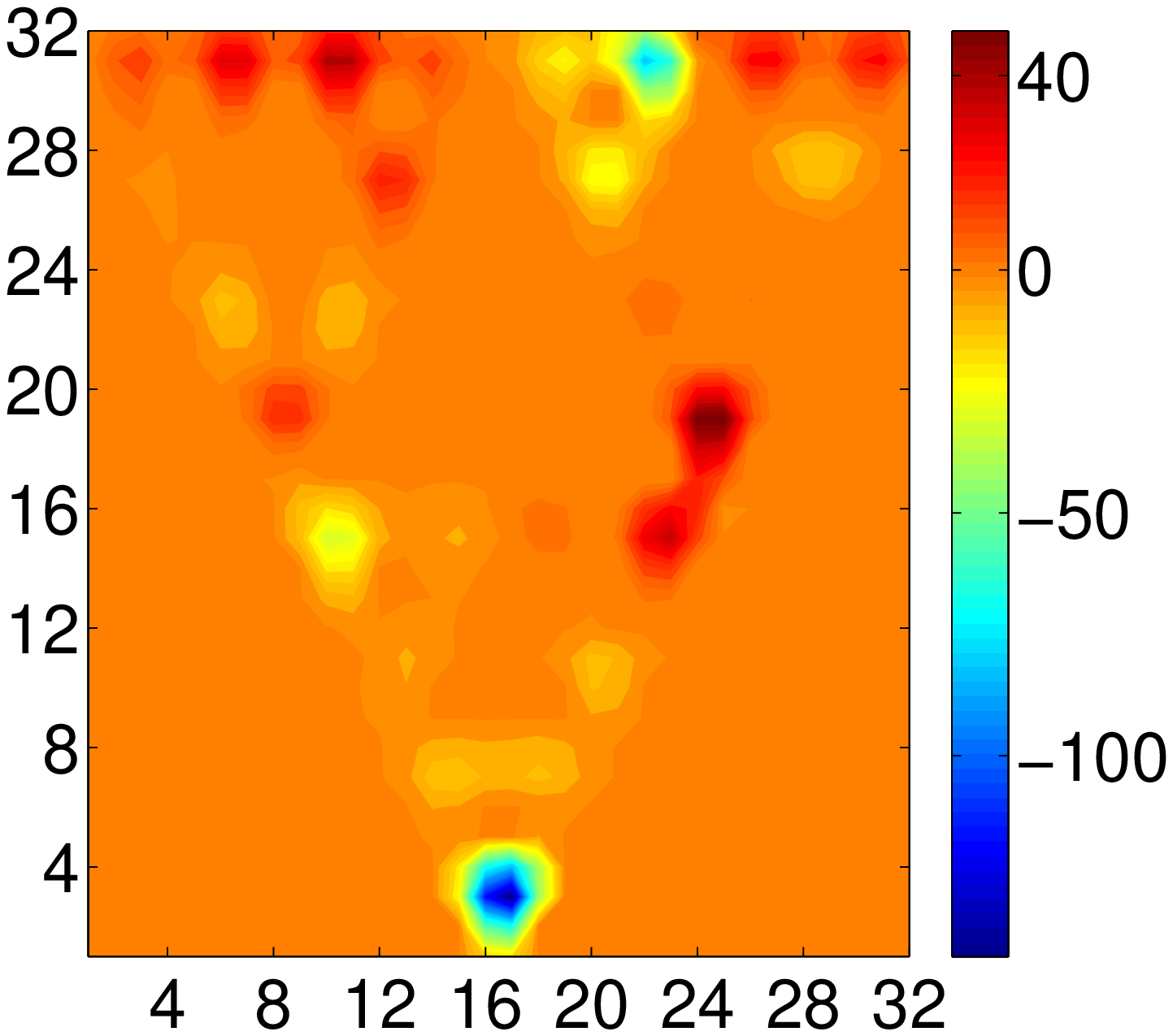}
	%\caption{Case 3}	
\end{minipage}
}
\caption{\raggedright (Color online) Local spin density and charge density under a strong bias  and $ E_{f}=-2.9 $ in fourth generation.}
\label{fig:sdqh:c4}
\end{figure}

The fourth generation is setting in the band bottom $ E_{b}=-4t_{0} $. Because reducing the sites of the sample and the width, $ 2a $, connecting with two leads, the Fermi-level moves to $ -2.9t_{0} $. Under the low bias, the spins and the charges transport hardly in the region of the upper line in Fig. \ref{fig:sdql:c4}. The self-similar feature of the fractal is not observed at all. Under high bias, there are accumulated charges and spins in the corners of building blocks of the sample in Fig. \ref{fig:sdqh:c4}. Similar selection of spin orientation was observed in this downward triangular SPG.

\subsection{Transmission}
\label{sec:tm}

\begin{figure}[!htb]
%\begin{flushleft}
\centering
\subfloat[\label{fig:tm:all}]{

%\begin{minipage}[h]{}
	%\centering
	\includegraphics[width=3.0in]{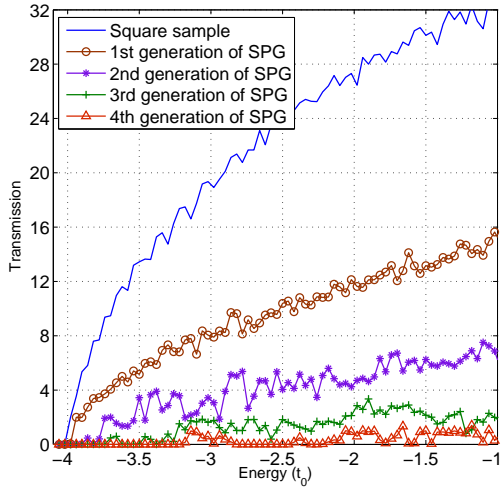}	
	%\caption{Case 2}
%\end{minipage}
} \\
\subfloat[\label{fig:tm:234}]{

%\begin{minipage}[h]{}
	%\centering
	\includegraphics[width=3.0in]{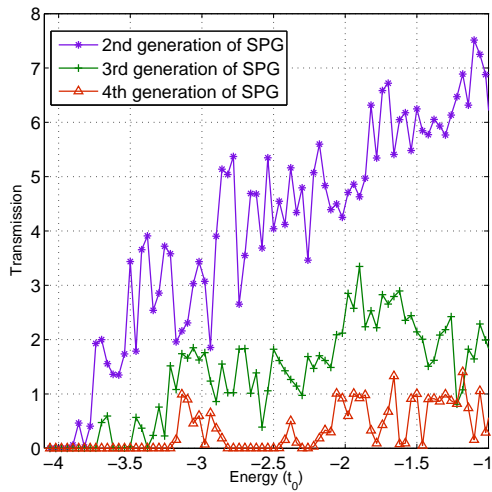}	
	%\caption{Zoom in on the case 2 (purple) , case 3 (green) and case 4 (red) from $ 0 $ to $ 8 $ in y-axis. Note that the central  triangle defect  are with effect in case 3 (green) and case 4 (red).}
%\end{minipage}
} 
\caption{\label{fig:tm} \raggedright (Color online) Transmission of SPG versus bias energy. The bias energy is normalized with the hopping energy $ t_0 $. (a) The transmission of a 2D square reference sample (blue line) starts from around $ -4t_0 $ and then increases with energy. The non-zero transmission of higher generation of SPG shifts to smaller values. (b) Zoom in on the higher generation of SPGs with the central defects and the transmission was suppressed significantly. Note that the threshold of higher generation of SPGs also shifts to left and the central  triangle defects  are with effect in third generation (green dagger-line) and fourth generation (red triangle-line).}

%\end{flushleft}
\end{figure}

In a 2D square sample, the transmission starts from $ -4t_0 $ shown as blue line in Fig. \ref{fig:tm:all}. For $ 1^{st} $ generation of SPG (brown) and $ 2^{nd} $ generation of SPG (purple), the transmissions are almost the same as in square sample (Fig. \ref{fig:tm:all}) but with curve slope change to linear dependence of energy. The lower threshold of transmission of $ 3^{rd} $ generation of SPG shifts from $ -4t_0 $ to $ -3.39t_0 $ in Fig. \ref{fig:tm:234}, and the transmission of $ 4^{th} $ generation shifts further to $ -2.9t_0 $. Our results indicated that the transmissions of fractal structure are affected both by the shape of the sample and the defects within the samples seriously. For $ 1^{st} $ generation of SPG, the leads contact the sample with $  $16 sites ($ 16a $) on both left and right. For $ 2^{nd} $ generation, the left and right leads contact the sample only with $ 8 $ sites ($ 8a $). Since the width of two leads are comparatively wider than the other higher generation of SPG, and thus the spin-polarized electrons can be through two leads smoothly. For $ 3^{rd} $ generation of SPG, the two leads are in contact with only $ 4 $ sites ($ 4a $) in each side and the transmission reduces heavily. It should be noted that the central defects of the higher generation of SPG make their transport behavior is topological different from the lower generation of SPG. The threshold of the transmission shifts from $ -4t_0 $ to $ -2.9t_0 $ with increasing number of defects, i.e., a higher generation of SPG. For the $ 4^{th} $ generation of SPG, a visible gap even appears within $ -2.7t_0 $ to $ -2.5t_0 $ (Fig. \ref{fig:tm:234}) and this arises from the self-similarity of SPGs. For the higher generation of SPG, the contact sites with leads reduces and the transmission also reduces accordingly for the narrowing of the conducting channel and most charges and spins are trapped at the bottom of fractal structure. A self-similarity behavior of transmission can be observed from $ 2^{nd} $ generation of SPG for the topological difference with the lower generation of SPG (Fig. \ref{fig:tm:234}).

\section{Conclusion}
In conclusion, we had studied spin transport in a planar Sierpinski gasket fractal formed by elementary triangle unit. We also compare the local charge density, the local spin density and also spin Hall effect on the planar Sierpinski gasket with a continuous planar and linear structure. For four different planar Sierpinski gasket fractal structures in our analysis, with a low bias, there are little fractal-like patterns in first generation and second generation. Moreover, because of too many holes on the third generation and fourth generation, the electrons become very localized and transport can be only observed along the linear edge. Therefore, hardly any fractal patterns were observed for both the local spin density and local charge density. Indeed there is quantum dot region was observed in first generation (Fig. \ref{fig:sdql:c1} and Fig. \ref{fig:sdqh:c1}) and also barely was observed in second generation (Fig. \ref{fig:sdqh:c2}). The behavior can be considered to be a similar arrangement of single level QD’s \cite{PhysRevB.65.245301, Chen:2008vn} sitting at the vertices of a continuous line. Also the spin density and the charge density indicate that the transmissions in fractal structures are smaller than linear structures. And this is consistent with the results of transmission. A very interesting phenomenon was observed in our studies, due to the fractal shape, spin transport can be separated from the charge transport and details mechanism still need to be studied (Fig. \ref{fig:szl:c3}). It should be noted that the asymmetry of a planar Sierpinski gasket fractal could trap the spin in the bottom from the spin Hall effect. The up and down spin-polarized electrons accumulate on the wide and narrow edges of a planar Sierpinski gasket fractal and thus the up and down spin electrons can then be selected from the asymmetrical shape. Under the appropriate condition, there is only one kind of spin electron can pass through the asymmetrical device. This geometrical selection provides a feasible way of producing pure spin current. 

The transmissions of four different generations are calculated numerically. As reducing the sites of the devices, the widths connection with two leads also decrease. From the analysis of different generation of SPG, it was found that the contact widths of the leads determine the upper bound of the transmission. The transmission affected by the connecting width with two leads and the threshold of the transmission depends on the number of defects.

\begin{acknowledgments}
We appreciate the Taiwan National Science Council Grant No. NSC 98-2112-M-002-012-MY3 for supporting of this work.
\end{acknowledgments}

%\bibliographystyle{apsrmp4-1.bst}
\bibliography{ref}

\end{document}